\documentclass{aa}
\usepackage{graphics,epsfig,txfonts}

\usepackage{natbib}
\bibpunct{(}{)}{;}{a}{}{,}

\newcommand{\oergcm}[1]{$10^{#1}$ erg cm$^{-2}$ s$^{-1}$}

\newcommand{\expo}[1]{$\times 10^{#1}$}

\newcommand{\kms}{km s$^{-1}$}
\newcommand{\nh}{N$_{\rm H}$}

\newcommand{\Halp}{H${\alpha}$}
\newcommand{\ltsima}{$\buildrel < \over \sim$}
\newcommand{\lsim}{\lower.5ex\hbox{\ltsima}}
\newcommand{\gtsima}{$\buildrel > \over \sim$}
\newcommand{\gsim}{\lower.5ex\hbox{\gtsima}}
\newcommand{\SN}{\hbox{SN\,1987\,A}}
\newcommand{\sn}{\hbox{SN\,1987\,A}}
\newcommand{\xmm}{XMM-Newton}

\begin{document}

\title{High Resolution X-Ray Spectroscopy of \object{SN\,1987\,A}: Monitoring with XMM-Newton}

%       \thanks{}

\author{R.~Sturm\inst{1} 
        \and F.~Haberl\inst{1}
        \and B.~Aschenbach\inst{1}
        \and G.~Hasinger\inst{1,2}
       }

\institute{Max-Planck-Institut f\"ur extraterrestrische Physik, Giessenbachstra{\ss}e, 85748 Garching, Germany
           \and Max-Planck-Institut f\"ur Plasmaphysik, Boltzmannstra{\ss}e 2, 85748 Garching, Germany
          }

\titlerunning{High Resolution X-Ray Spectroscopy of \object{SN\,1987\,A}: Monitoring with XMM-Newton}
\authorrunning{Sturm et al.}
 
\offprints{R. Sturm, \email{rsturm@mpe.mpg.de}}
 
\date{Received 18 September 2009 / Accepted 1 February 2010}

\abstract{We report the results of our \xmm\ monitoring of \sn. 
          The ongoing propagation of the supernova blast wave through the inner circumstellar ring 
          caused a drastic increase in X-ray luminosity during the last years, 
          enabling detailed high resolution X-ray spectroscopy with the Reflection Grating Spectrometer.} 
         {The observations can be used to follow the detailed evolution of the arising supernova remnant.}
	 {The fluxes and broadening of the numerous emission lines seen in the dispersed spectra 
          provide information on the evolution of the X-ray emitting plasma and its dynamics.
          These were analyzed in combination with the EPIC-pn spectra, 
          which allow a precise determination of the higher temperature plasma. 
          We modeled individual emission lines and  
          fitted plasma emission models.} 
         {Especially from the observations between 2003 and 2007 we can see a significant evolution of the plasma parameters 
          and a deceleration of the radial velocity of the lower temperature plasma regions. 
          We found an indication (3$\sigma$-level) of an iron K feature in the co-added EPIC-pn spectra.}
         {The comparison with Chandra grating observations in 2004 yields a clear temporal coherence of the spectral evolution and the 
          sudden deceleration of the expansion velocity seen in X-ray images $\sim$6100 days after the explosion.}

 
\maketitle

\section{Introduction}
The circumstellar ring system around \SN\ was ejected by the progenitor star approximately 20\,000 years before the supernova explosion.
About 10 years after the explosion the blast wave started to propagate through the inner ring, 
causing compression, heating and ionization of its matter.
At that time bright unresolved regions, so called hot spots, 
appeared all around the ring in HST images \citep{2000ApJ...537L.123L} 
and the rather linear increase in the soft X-ray band, 
as observed with ROSAT since 1992 \citep{1994A&A...281L..45B,1996A&A...312L...9H}, 
was followed by an exponential brightening as monitored with \xmm, Chandra, Suzaku and Swift. 
Recently, some flattening of the flux increase is seen \citep[][and references therein for the individual fluxes]{2009PASJ...61..895S}.

The X-ray spectra are interpreted as thermal emission composed of a lower temperature component ($\sim$0.5~keV) 
and a higher temperature component ($\sim$2.5~keV).
Deep Chandra grating observations allowed the measuring of the bulk gas velocity due to spectral line deformation.
Surprisingly these values were lower than the velocities expected from the plasma temperatures.
This suggests the contribution of reflected shocks, 
additional to the 'normal' forward shock \citep{2005ApJ...628L.127Z,2006ApJ...645..293Z,2009ApJ...692.1190Z,2008ApJ...676L.131D}. 
The expansion velocity derived from Chandra X-ray images is even higher. About 6100 days after the explosion
a sharp deceleration to 1600$-$2000~\kms\ was observed \citep{2009ApJ...703.1752R}.

Shocks transmitted into denser regions of the ring have a slower shock wave velocity 
and therefore can be responsible for the low temperature component.
This interpretation is supported by the morphology seen in optical and X-ray images \citep{2007AIPC..937....3M}, 
as well as the similar evolution of the soft X-ray flux (0.5-2.0~keV) 
and the evolution of highly ionized optical emission lines from the hot spots \citep{2006A&A...456..581G}. 
This emission might be caused by even slower radiative shocks.

Renewed radio emission of \SN\ was detected in July 1990 \citep{1990IAUC.5086....2T}.
A continuously rising flux and increasing source radius has been observed by the ATCA since then \citep{2007AIPC..937...86G,2008ApJ...684..481N}.
The increase of the radio light curve matches the evolution of the hard X-ray flux (2-10~keV) quite closely \citep{2005ApJ...634L..73P,2007AIPC..937...33A}, 
thus the synchrotron radio emission may originate in the hot thermal plasma between the forward and reverse shock
and a fraction of the hard X-ray flux may also have a non-thermal origin.
Broad \Halp\ and Ly$\alpha$ lines suggest the presence of the reverse shocks \citep{2003ApJ...593..809M,2006ApJ...644..959H}.

This study reports on the yearly \xmm\ monitoring observations of \SN\ between January 2007 and January 2009 
together with one prior observation from May 2003, 
making it possible to reveal the detailed plasma evolution. 
We measured light curves and widths of individual emission lines and 
analyzed the spectral plasma evolution by fitting thermal plasma emission models.

\section{Observations and Data Reduction}

%------------------------------------------------------------------------------------------------
\begin{table*}
\caption{\xmm\ observations of \sn\ and data selection.}
\label{tab:xray-obs}
\begin{center}
\begin{tabular}{llccccrr}
\hline\hline\noalign{\smallskip}
\multicolumn{1}{c}{Date} &	
\multicolumn{1}{c}{Time} &
\multicolumn{1}{c}{Satellite } &
\multicolumn{1}{l}{Instrument} &	
\multicolumn{1}{c}{Read-out} &
\multicolumn{1}{c}{Filter} &
\multicolumn{1}{c}{Net Exp} &   
\multicolumn{1}{r}{Counts$^{(a)}$} \\
\multicolumn{1}{l}{} &	
\multicolumn{1}{c}{(UT)} &
\multicolumn{1}{c}{Revolution} &
\multicolumn{1}{c}{} &
\multicolumn{1}{c}{Mode} &
\multicolumn{1}{c}{} &
\multicolumn{1}{c}{[s]} &   
\multicolumn{1}{r}{} \\
\noalign{\smallskip}\hline\noalign{\smallskip}
  2003 May 10-11  &    16:41-23:37     &  0626  & EPIC-pn &	  SW, 6ms & medium &       34983   &  13712   \\
  		  &    11:42-23:38     & 	& RGS1    &	  Spectro & -	   &       112164  &  3061    \\
  		  &    10:47-23:38     & 	& RGS2    &	  Spectro & -	   &       111671  &  3635    \\
\noalign{\smallskip}\hline\noalign{\smallskip}
  2007 Jan 17-19  &    19:28-01:18     &  1302  & EPIC-pn &	  FF, 73ms& medium &       72478   &  117055  \\
  		  &    18:23-01:19     & 	& RGS1    &	  Spectro & -	   &       110528  &  7797    \\
  		  &    18:23-01:18     & 	& RGS2    &	  Spectro & -	   &       110475  &  11989   \\
\noalign{\smallskip}\hline\noalign{\smallskip}
  2008 Jan 11-13  &    21:32-04:07     &  1482  & EPIC-pn &	  FF, 73ms& medium &       79549   &  166180  \\
  		  &    20:30-04:14     & 	& RGS1    &	  Spectro & -	   &       113617  &  10022   \\
  		  &    20:30-04:14     & 	& RGS2    &	  Spectro & -	   &       113846  &  15921   \\
\noalign{\smallskip}\hline\noalign{\smallskip}
  2009 Jan 30-31  &    18:40-22:27     &  1675  & EPIC-pn &	  FF, 73ms& medium &       72925   &  185071  \\
  		  &    18:17-22:36     & 	& RGS1    &	  Spectro & -	   &       101312  &  11528   \\
  		  &    18:17-22:36     & 	& RGS2    &	  Spectro & -	   &       101547  &  17346   \\
\noalign{\smallskip}\hline\noalign{\smallskip}
\end{tabular}
\end{center}
$^{(a)}$Net. counts in the $0.53-10.0$~keV (EPIC-pn) and 0.35$-$2.1~keV (RGS)  band\\
\end{table*}
%------------------------------------------------------------------------------------------------

%------------------------------------------------------------------------------------------------
\begin{figure*}
%\sidecaption
 \resizebox{0.95\hsize}{!}{
 \includegraphics[width=6cm,angle=-90]{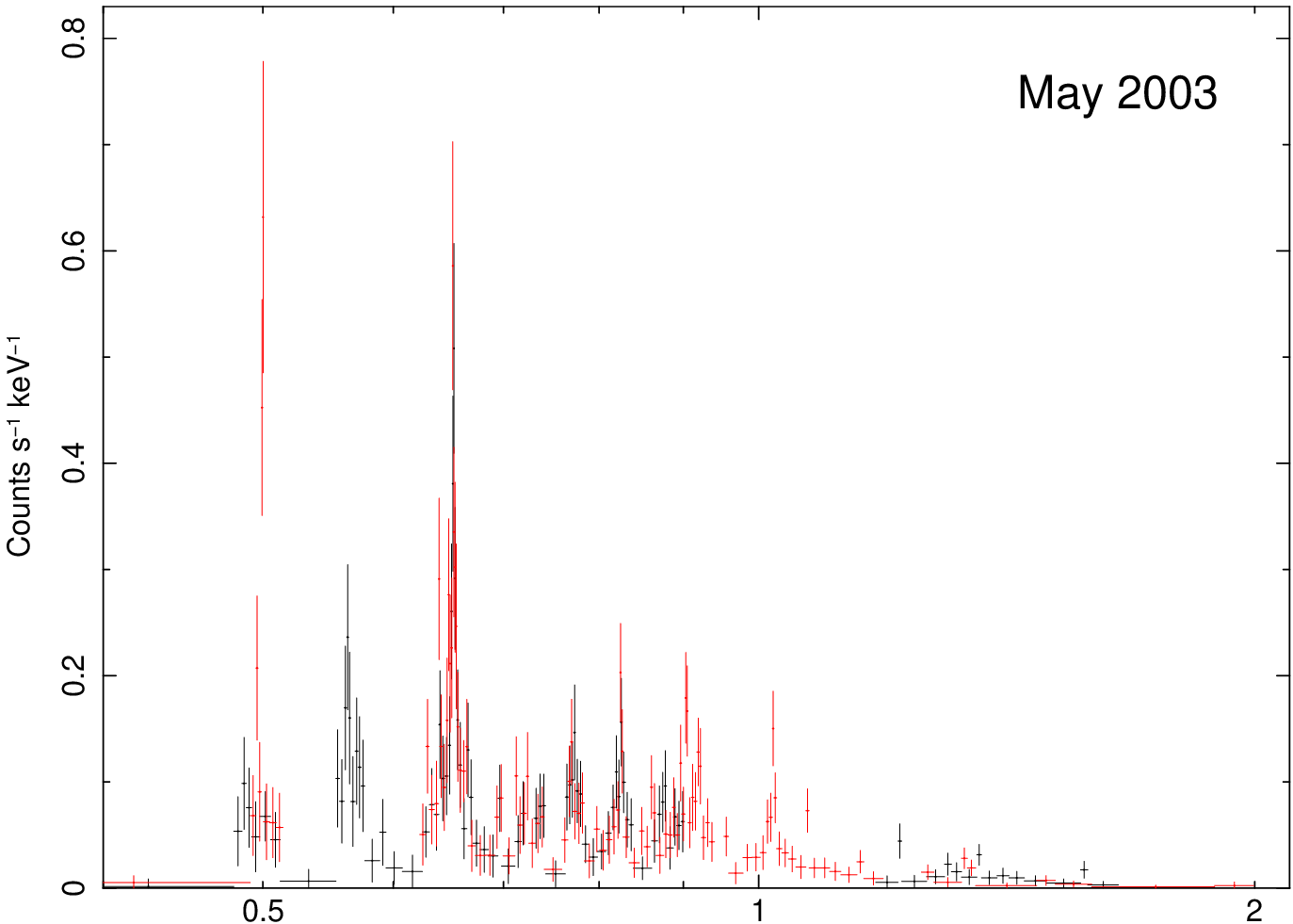}\quad
 \includegraphics[width=6cm,angle=-90]{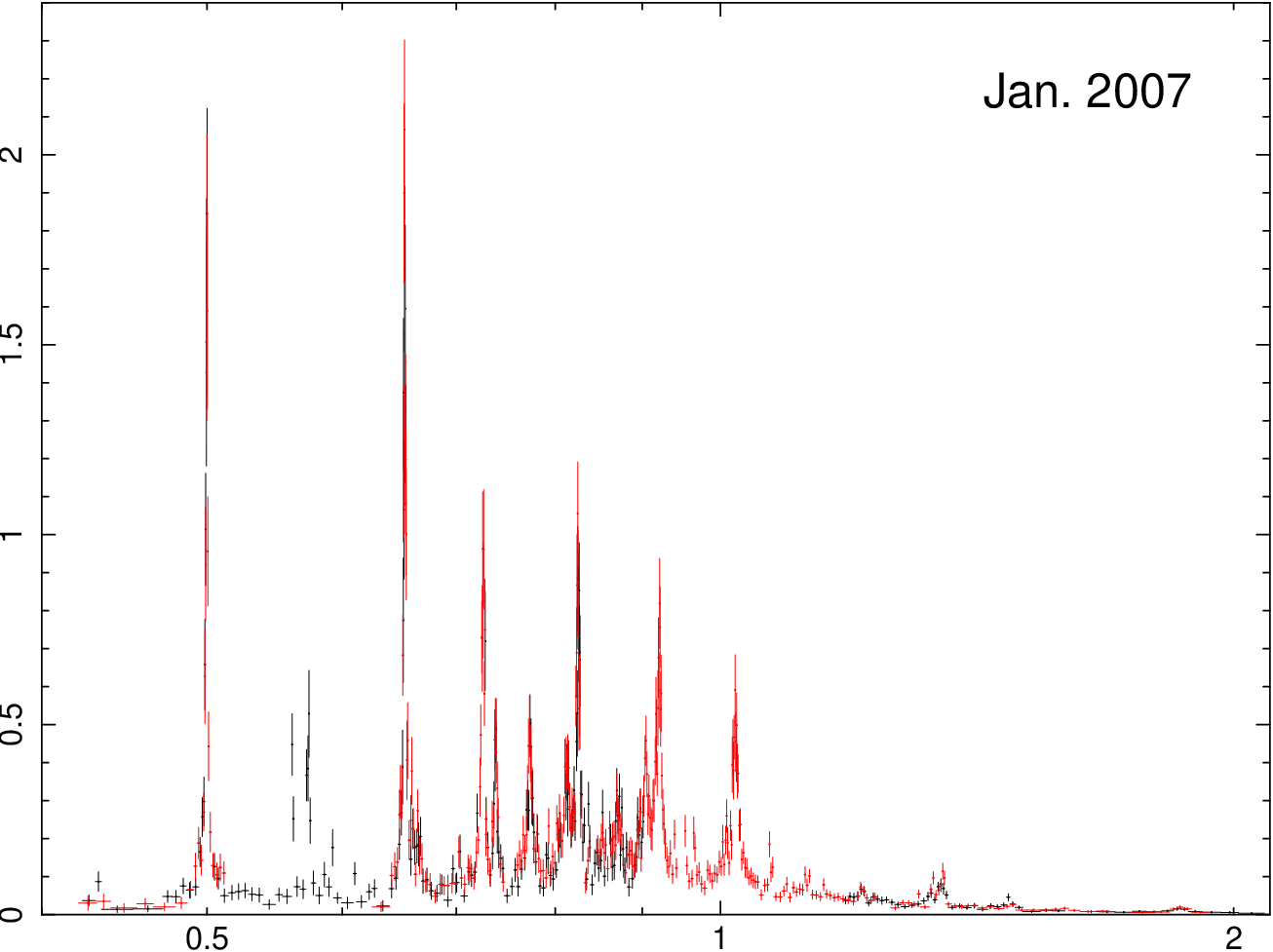}
}
\newline 
 \resizebox{0.95\hsize}{!}{
 \includegraphics[width=6cm,angle=-90]{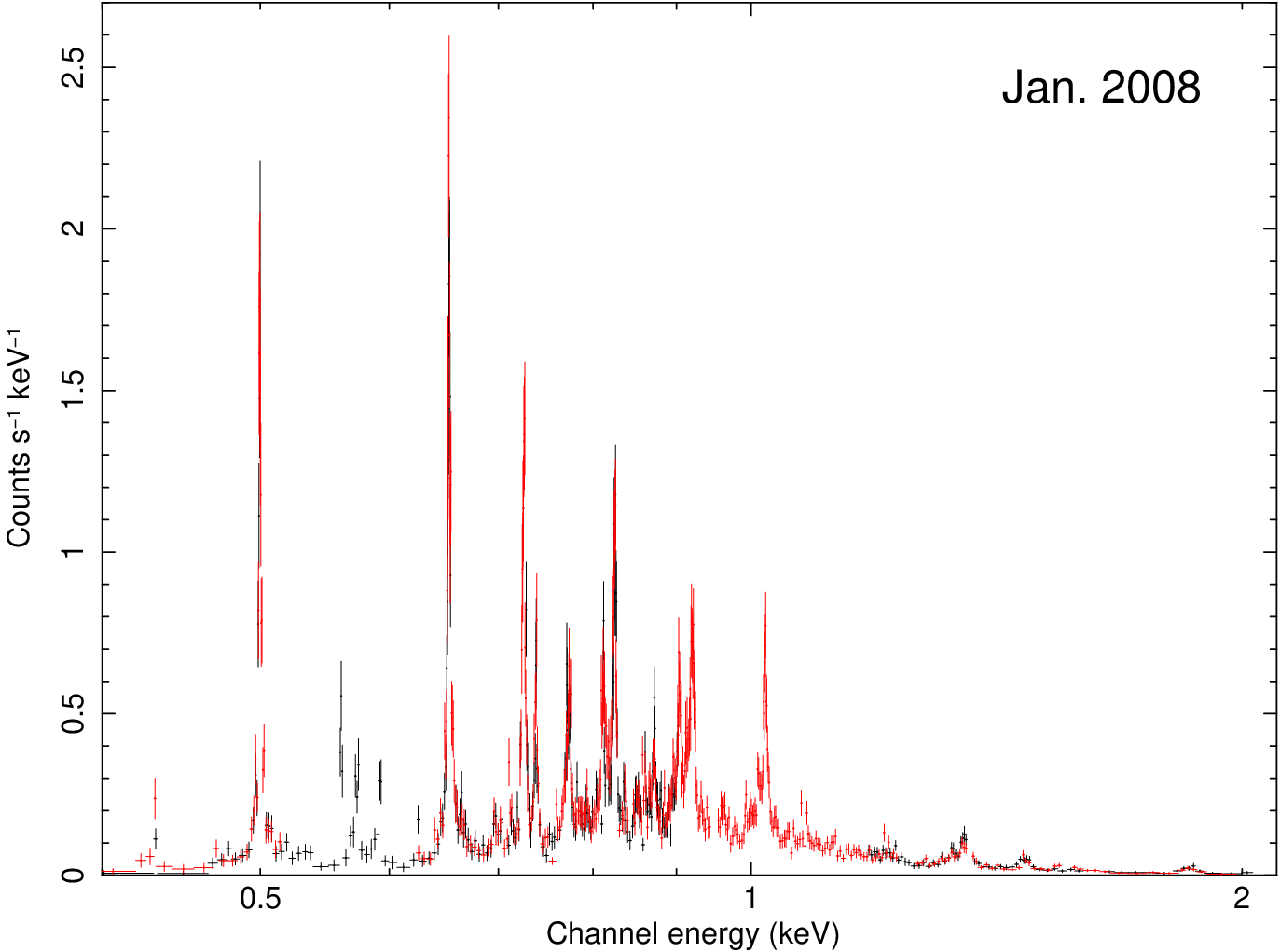}\quad
 \includegraphics[width=6cm,angle=-90]{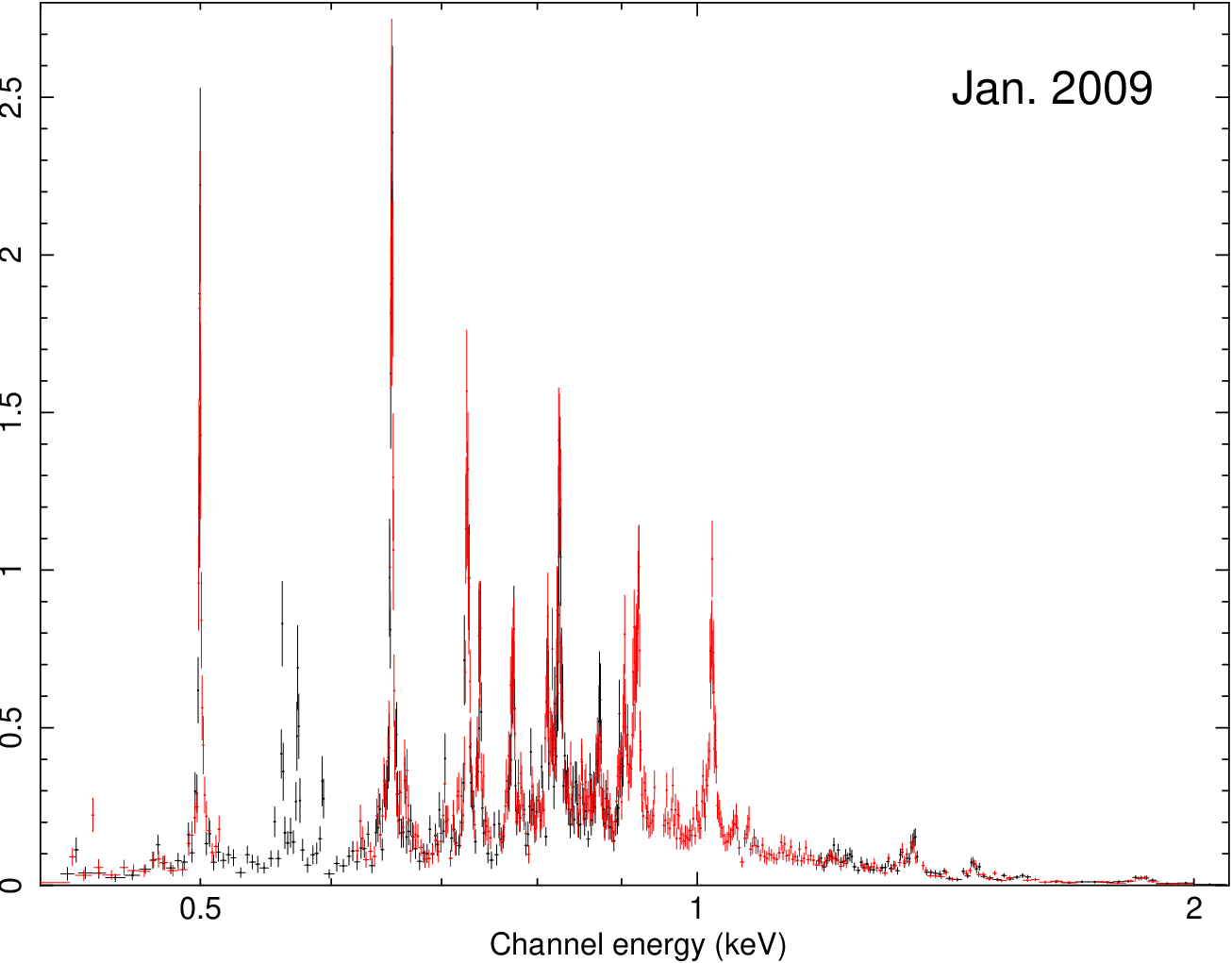}
}

\caption[]{
    The RGS1 (black) and RGS2 (red) spectra of \SN\ observations in 2003, 2007, 2008 and 2009. 
    A clear evolution in emission line ratios is seen. E.g. compare the nearly absent \ion{Fe}{xvii} lines at $\sim$$0.72~keV$ in 2003 
    to the strong lines seen in the later observations. Also the \ion{Ne}{x} Ly$_{\alpha}$ (1.02~keV) flux increases relative to the \ion{Ne}{ix} triplet ($\sim$0.91~keV).
   }
   \label{fig:spectra}
   
\end{figure*}
%------------------------------------------------------------------------------------------------

%observations
Since January 2007 we have performed a yearly monitoring of \SN\ using \xmm\ \citep[][]{2001A&A...365L...1J}. 
Including an earlier observation from May 2003, four observations with an exposure each exceeding 100~ks were accomplished. 
A summary is listed in Table~\ref{tab:xray-obs}. Earlier observations and the May 2003 data were already analyzed by 
\citet{2006A&A...460..811H}, and \citet{2008ApJ...676..361H} evaluated the Jan. 2007 observation concentrating on the elemental abundances.
In this study we use consistent models to analyze the two new observations in combination with the previous data, 
to obtain the evolution of the X-ray spectra as seen by the
Reflection Grating Spectrometer \citep[RGS,][]{2001A&A...365L...7D}.
To extend the X-ray band up to 10~keV, we included also the EPIC-pn \citep[][]{2001A&A...365L..18S} spectra in our analysis.

%reduction
We used XMM-Newton SAS 8.0.0\footnote{Science Analysis Software (SAS), http://xmm.vilspa.esa.es/sas/} to process the data. 
For the extraction of EPIC-pn spectra, single-pixel events in good time intervals (GTIs) 
with low background (threshold at 8 cts\,s$^{-1}$\,arcmin$^{-2}$) were selected from the source region, centered on \SN, 
and a point source free background region, each with a radius of 30\arcsec.
The RGS spectra were obtained  using {\tt rgsproc}, 
and GTIs (RATE $<$ 2.0) were used to select low background intervals.
In the 2003 observation, the Honeycomb nebula is located on the dispersion axis. 
Using the intrinsic energy resolution of the RGS Focal Plane Camera CCDs, counts from the Honbycomb nebula
are excluded.
For the other three observations, the Honeycomb nebula lies at the cross-dispersion axis at the 
very rim of the CCDs. By comparing our background spectra with the RGS background model 
(created by {\tt rgsbkgmodel}), we find no significant contribution from the 
Honeycomb nebula to the background spectra.

%co-added RGS
To obtain a detailed spectrum with $\sim$438~ks exposure, the RGS spectra of the four observations 
were added with {\tt rgscombine}, which also calculates combined response files.
To allow proper application of the $\chi$-statistics, 
all spectra are binned to have at least 30 (RGS) or 20 (EPIC-pn) counts per bin.
Due to the high statistics of the RGS-spectra a binning of 30 cts/bin
does not influence most emission lines which have far more than 30 cts
in most bins. thus there is no influence on the measured line parameters.

\section{Spectral Analysis}

Spectral fitting was performed using {\tt XSPEC} \citep{1996ASPC..101...17A} version 12.5.0x.
The errors are given for certain $\Delta \chi^2$ ranges.
Generally, $\Delta \chi^2=2.71$ is associated with the 90\% 
confidence range for one parameter of interest.

\subsection{Identification of Emission Lines}

%--------------------------
\begin{figure*}
\begin{center}
  \resizebox{0.95\hsize}{!}{\includegraphics[clip=,angle=-90]{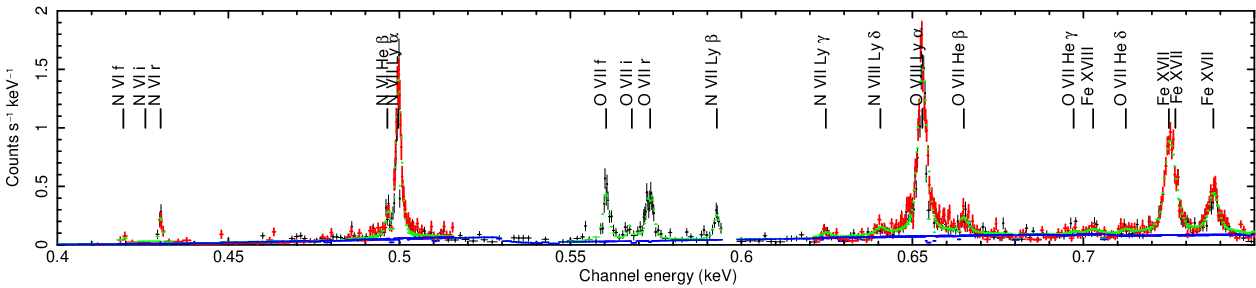}}
  \resizebox{0.95\hsize}{!}{\includegraphics[clip=,angle=-90]{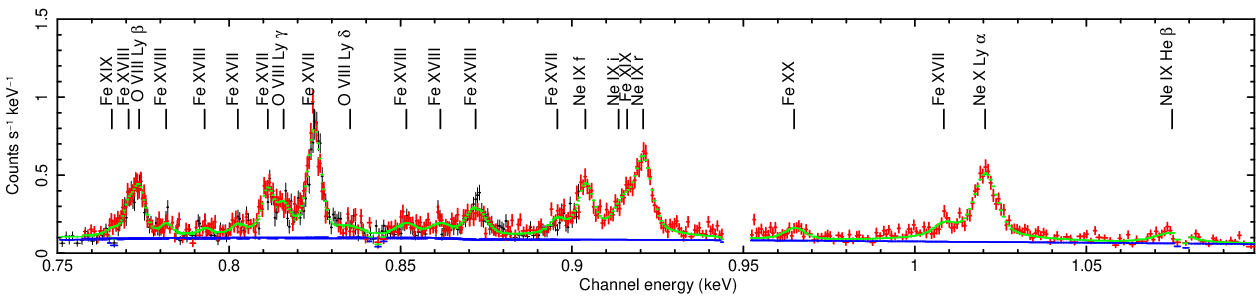}}
  \resizebox{0.95\hsize}{!}{\includegraphics[clip=,angle=-90]{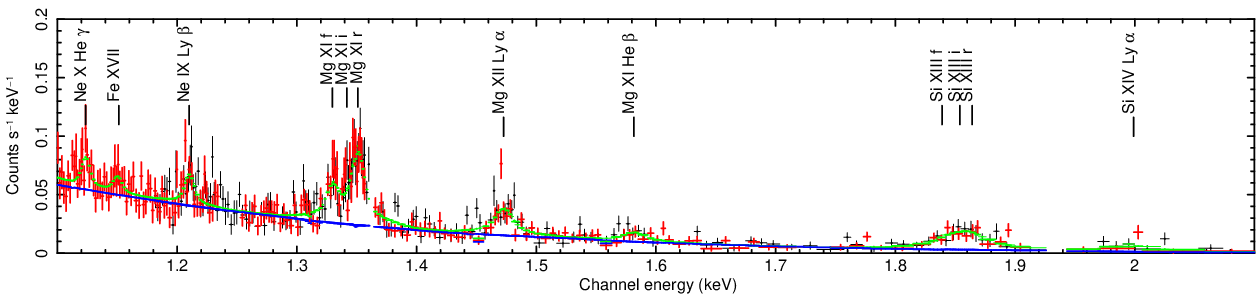}}

\caption[]{Co-added RGS1 (black) and RGS2 (red) spectra from all 4 observations of \SN\ 
           together with the best fit empirical model (green) and the absorbed bremsstrahlung continuum (blue).
           The line labels mark the strongest lines as expected from the two component {\tt vpshock} model 
           at the best fit energy of the line center. 
}
\label{fig:sumspec}
\end{center}
\end{figure*}
%--------------------------

%sumspec
Since the RGS spectra of \SN\ are dominated by emission lines (see Fig. \ref{fig:spectra}), 
we first identified the individual lines using the co-added spectra. 
We constructed an empirical model and fitted it to the combined RGS1 and RGS2 spectra simultaneously.

\begin{figure}
  \resizebox{\hsize}{!}{\includegraphics[angle=-90]{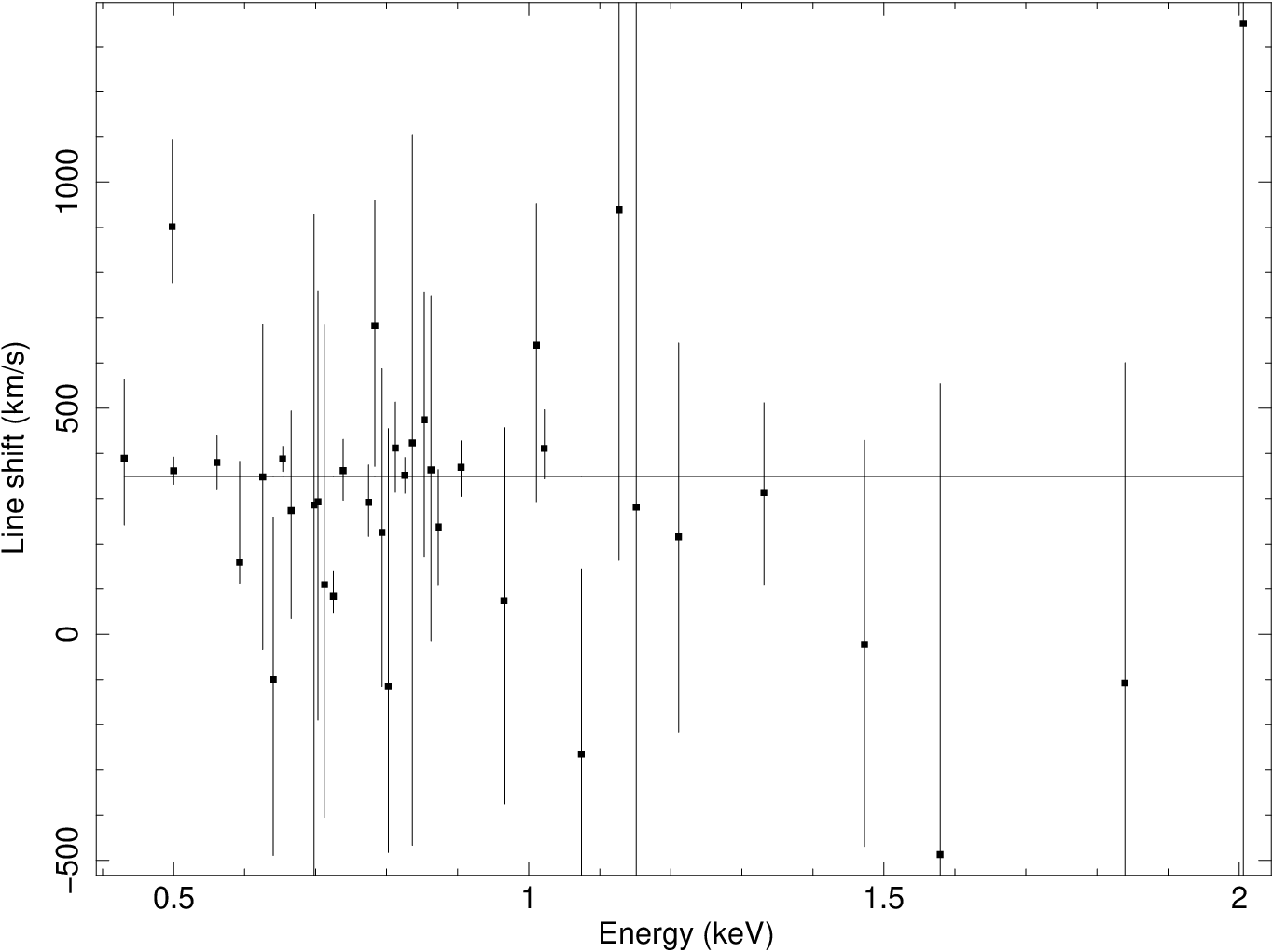}}
  \caption{Line shifts according to the empirical model. The horizontal line shows the best fit for a systemic redshift.
 Errors are for $\Delta\chi^2=2.71$ confidence range.}
  \label{fig:lineshift}
\end{figure}

As quasi-continuum we used a thermal bremsstrahlung model with absorption, 
which is formally a good approximation to the observed continuum, 
containing also radiative recombination and two photon decays.
This component also represents various weak emission lines, 
which are not resolved in the RGS spectra. 
A constant factor was allowed to vary between the two RGS spectra. 
We first fitted the strong lines with Gaussian profiles. 
The high statistical quality of the summed spectrum required 
the introduction of further weak lines where we noticed residua. 
Our final model comprises 53 lines as listed in Table~\ref{tab:lines}. 
Fitted to the co-added spectra, all lines have a normalization inconsistent with 0 for $\Delta \chi^2 = 2.71$ 
(except for the intercombination lines of helium-like N and Mg).
For line complexes the line energies were combined to one value according to their ratio 
in ATOMDB 1.3.1\footnote{http://cxc.harvard.edu/atomdb/} 
and the line widths were linked. For weak lines it was not possible to fit individual line widths, 
thus we coupled them with the next strong line.
We get a best fit with $\chi^2/{\rm dof} = 1843/1465$. 
The summed RGS spectra are shown in Fig.~\ref{fig:sumspec} together with the best fit model
and the bremsstrahlung quasi-continuum (kT$ = 402_{-6}^{+5}$ eV).
The line labels mark the best fit energy of the fitted Gaussian and name the strongest emission line expected 
from plane-parallel shocked plasma emission codes (see below) at this energy. 
With these identifications we obtain line shifts 
consistent with a systemic redshift of $349\pm24$ \kms (see Fig.~\ref{fig:lineshift}).

%--------------------------
%--------------------------

%--------------------------
\begin{figure*}
  \resizebox{1.0\hsize}{!}{\includegraphics[clip=,width=6.0cm,angle=-90]{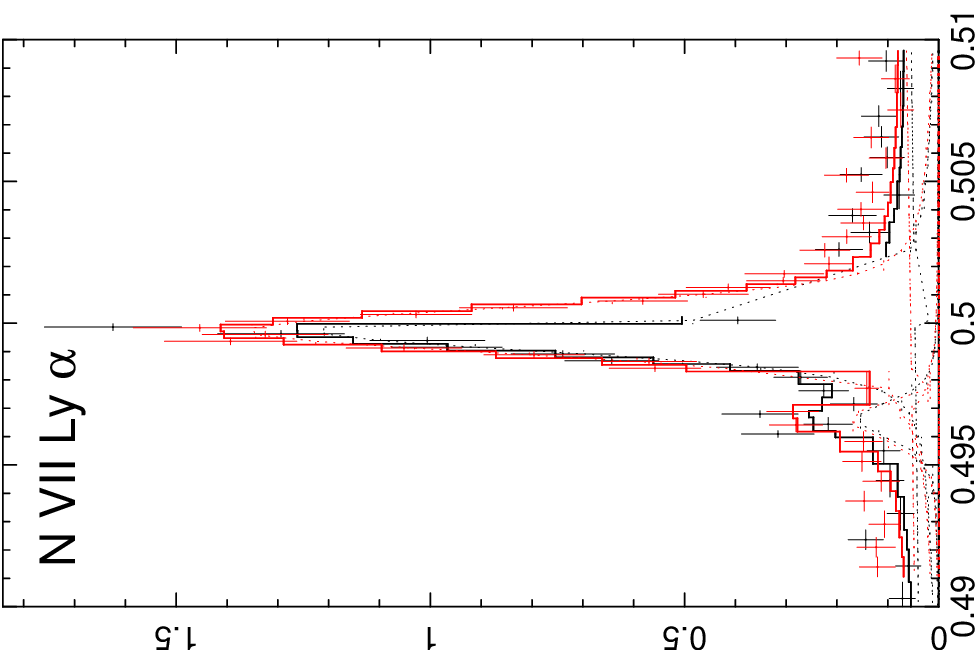}
                           \includegraphics[clip=,width=6.0cm,angle=-90]{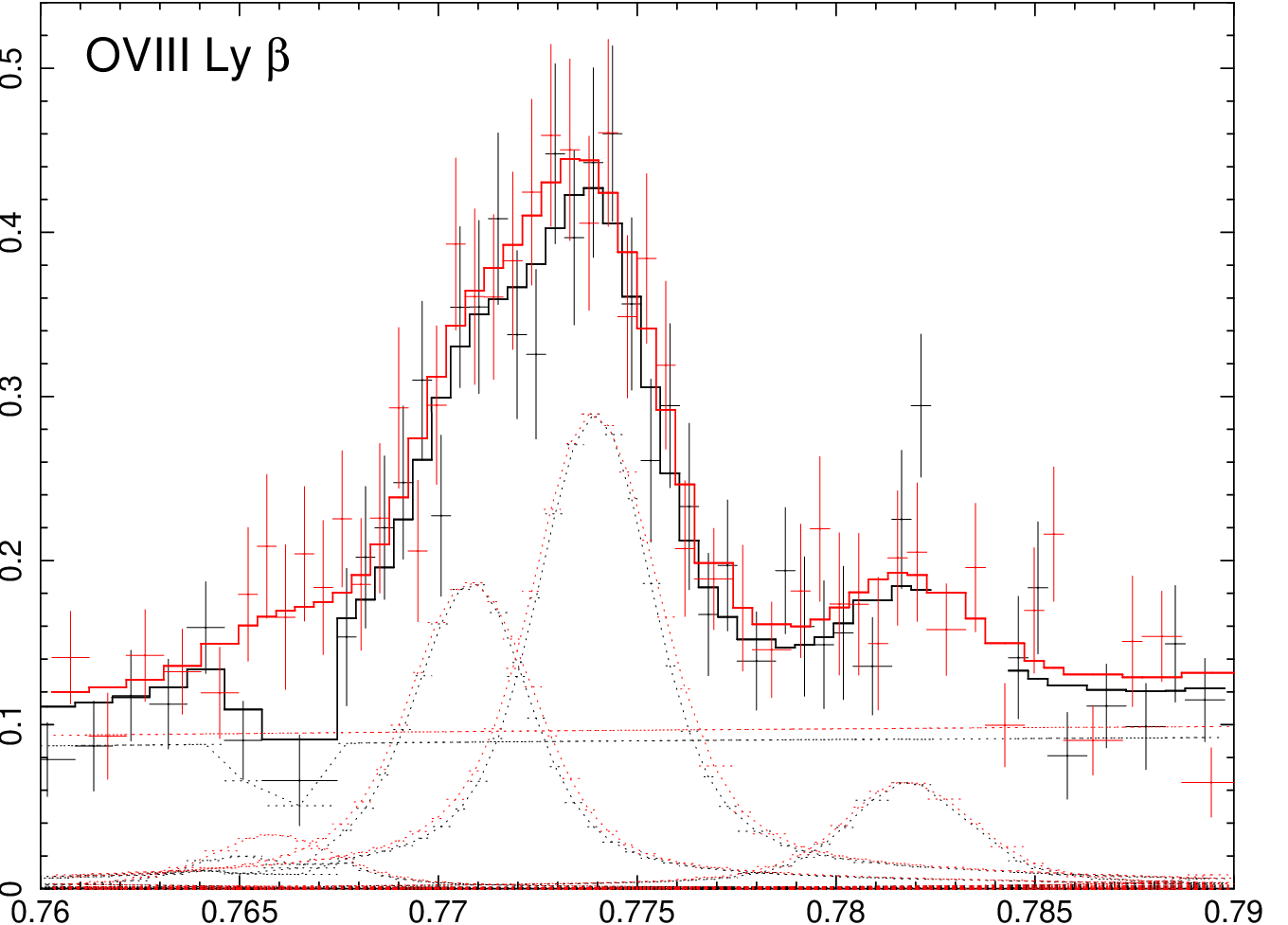}
                           \includegraphics[clip=,width=6.0cm,angle=-90]{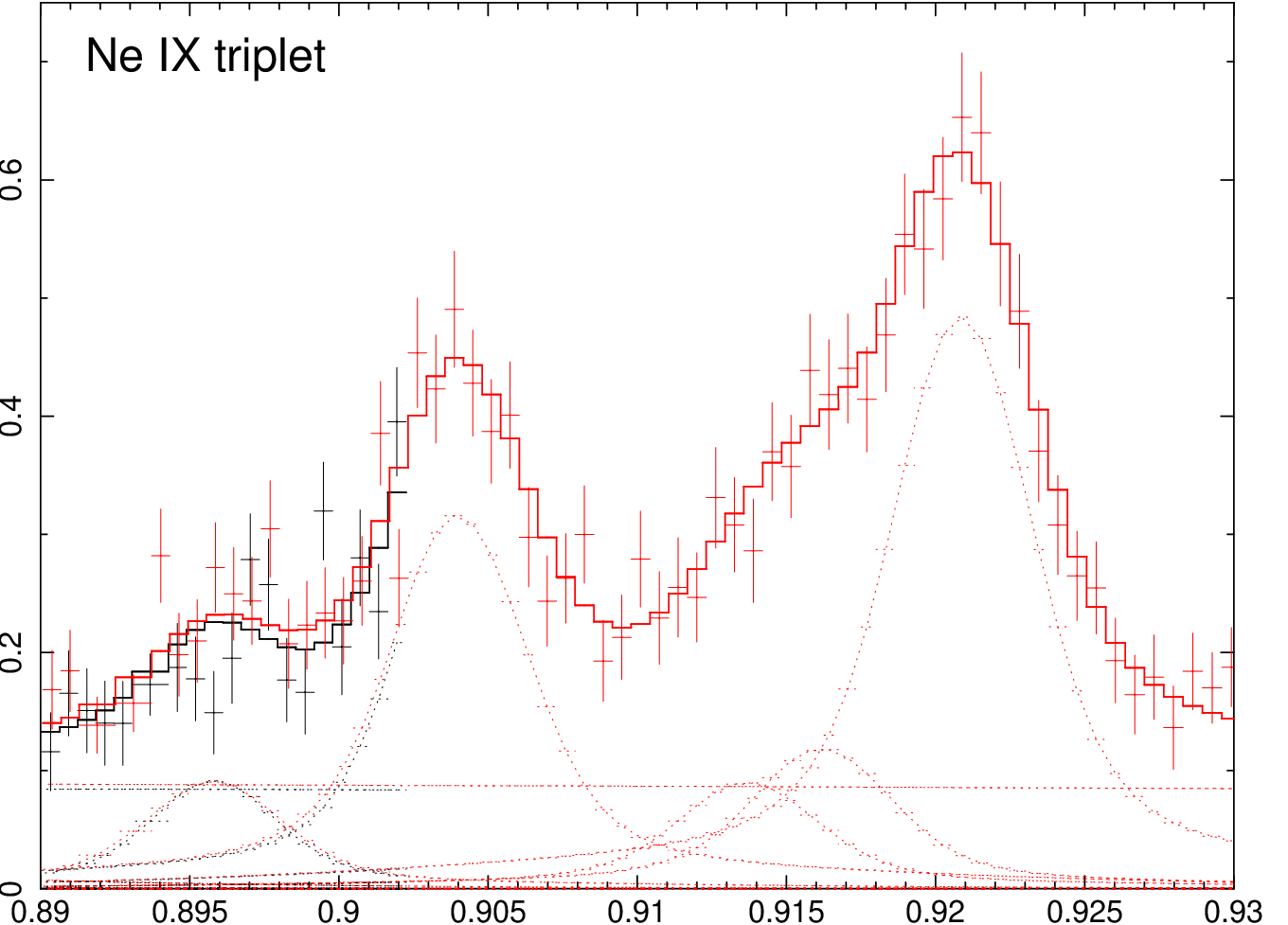}
                           \includegraphics[clip=,width=6.0cm,angle=-90]{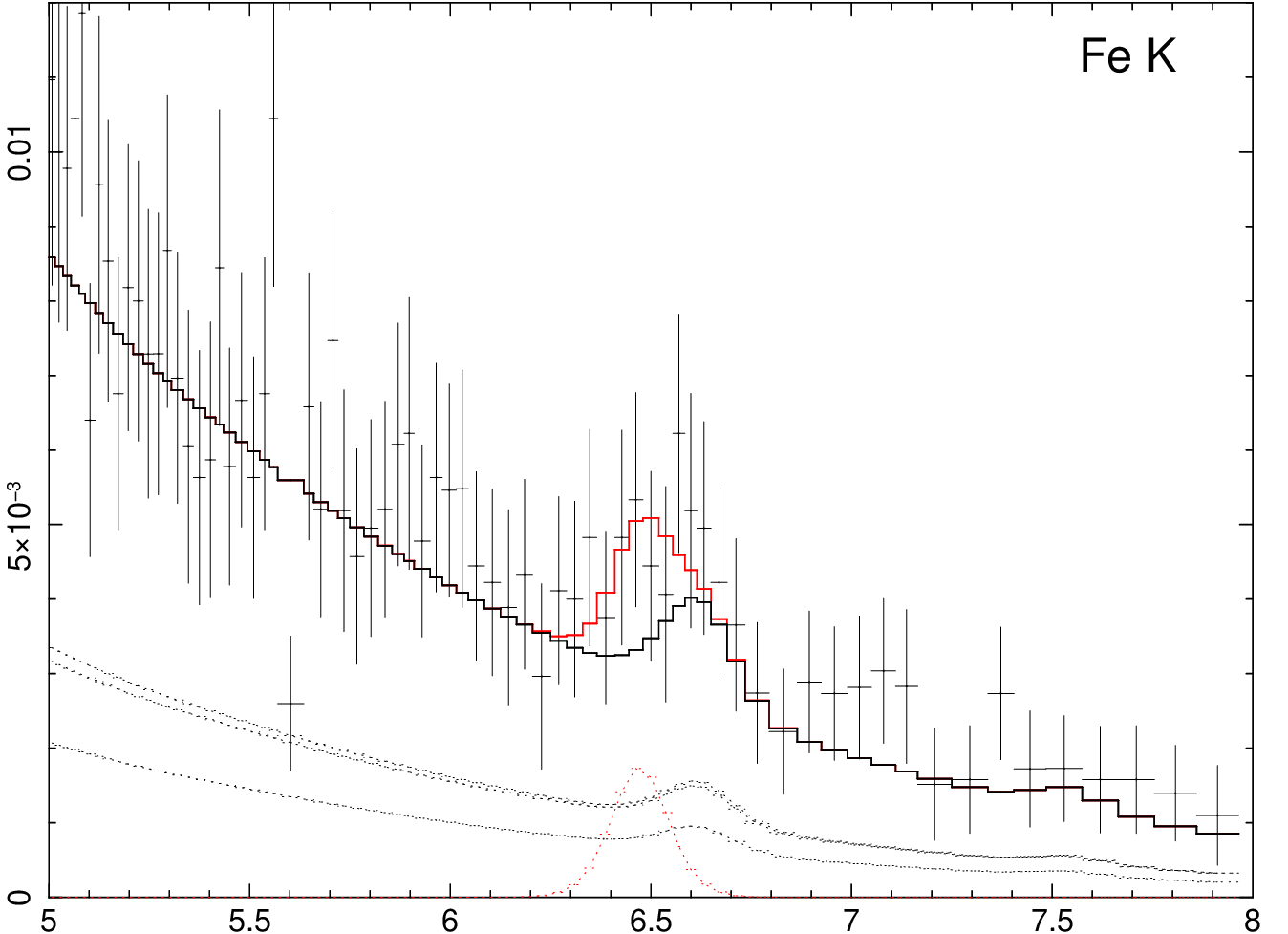}}
 \caption{Examples for emission line complexes from the summed spectra: 
          \ion{N}{vii} Ly $_{\alpha}$, \ion{O}{viii} Ly$_{\beta}$ and the \ion{Ne}{ix} triplet 
          as seen by the RGS1 (black) and RGS2 (red). Also the empiric model and its components are plotted.
          In the right panel an indication for a Fe~K feature is seen in the summed EPIC-pn spectrum of the newer three observations.
          A sum of the high temperature component plane-parallel shock models derived from the individual observations (black) 
          is compared with a model containing an additional Gaussian (red). Horizontal axes: Channel energy in keV, Vertical axes: Counts s$^{-1}$ keV$^{-1}$.
         }
\label{fig:lines}
\end{figure*}
%--------------------------

%-------comments on individual lines
The high statistical quality of the co-added RGS spectra allowed us to investigate emission lines in unprecedented detail.
%N VI He 
E.g. the \ion{N}{vi} helium-like triplet at $\sim$0.42~keV is clearly seen in the summed spectra (cf. Fig.~\ref{fig:sumspec}).
%N VI He beta
Other examples of emission line complexes are shown in Fig.~\ref{fig:lines}:
Next to the \ion{N}{vii} Ly$_{\alpha}$ (at 0.50~keV) line we found an additional line at slightly lower energy (see Fig.~\ref{fig:lines}), 
most likely \ion{N}{vi} He$_{\beta}$, which is the strongest line at this position in the plasma codes. 
But also argon lines contribute here which may cause the higher line shift. 
In previous analyses with lower statistics this line likely was blended with the $\sim$10 times more luminous \ion{N}{viii} Ly$_{\alpha}$ line.
%O VIII Ly$_{\beta}$
The line shape at the energy of the \ion{O}{viii} Ly$_{\beta}$ (0.77~keV) line is inconsistent with a Gaussian profile 
and the line width for a single Gaussian is outstandingly higher than found for the surrounding lines (2~eV vs. 0.25~eV). 
Plasma models suggest two iron lines at slightly lower energies, that were included in our model.
%Ne triplet
The shape of the \ion{Ne}{ix} triplet ($\sim$0.91~keV) is also not reproducible by three Gaussians. Here \ion{Fe}{xix} (917.1 eV) may contribute to the flux, 
although this line is not expected to be strong from emission codes.

%sigma
For strong lines in the summed spectra, we find measured line widths inconsistent with zero 
within a confidence range of $\Delta\chi^2=2.71$, 
demonstrating that line broadening can be detected significantly with RGS. 
Also a trend of increasing width with rising energy is seen. 
A fit of an energy dependent line width $\sigma(E)= \sigma_0 (E/1{\rm keV})^{\alpha}$ yields 
$\sigma_0=(1.35\pm0.36)$~eV and $\alpha=2.05_{-0.72}^{+0.66}$. 
The power law index agrees well with the Chandra LEGT 2007 results \citep{2009ApJ...692.1190Z}.

In a second step we fitted the empirical model to the spectra of the individual observations 
to follow the evolution of the individual emission lines in a consistent way. Only parameters 
with expected time dependence, i.e. the line fluxes and the parameters of the bremsstrahlung 
continuum, were allowed to vary. The absorption as well as the center energy and width of the Gaussians 
were fixed at the values obtained from the summed spectra. Here these values can be determined 
more precisely than in the individual spectra. 
Using the RGS spectra with the best statistics (i.e. the 2009 data), we compared line widths 
fixed in the individual fits with those allowing line widths free in the fits.
We found that 20 of 21 line widths are consistent. % (with respect of 1 line (\ion{O}{viii} Ly$_{\alpha}$)). 
A time dependent analysis of the line widths is done in section~\ref{cap:analysis:plasma}, 
by using plasma models and assuming a power law dependence.
The line fluxes with fixed and variable widths also are consistent and we conclude that the fixed line widths do not influence the derived line fluxes.

\begin{figure*}
  \resizebox{1.0\hsize}{!}{\includegraphics[clip=,width=6cm,angle=-90]{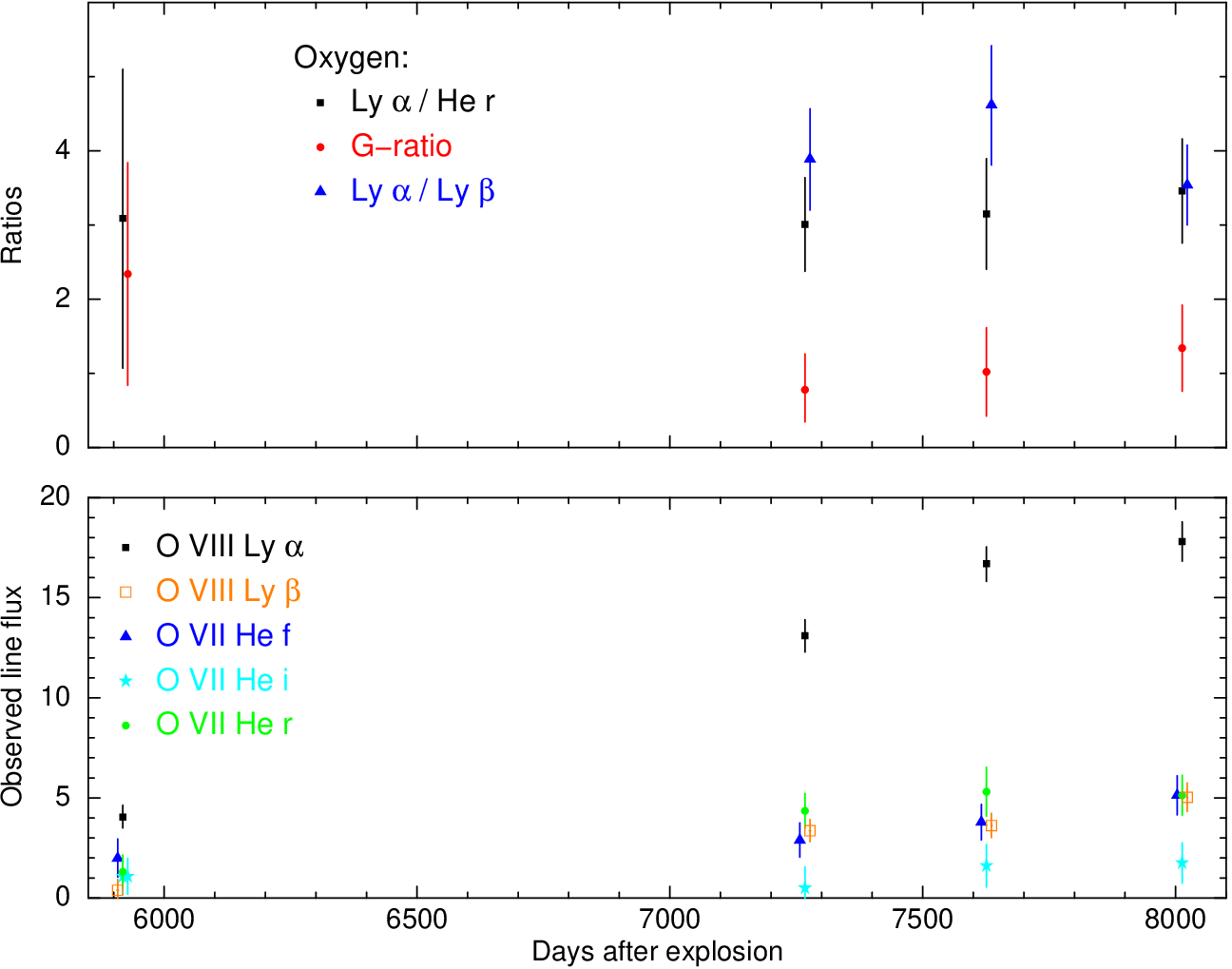} \qquad
                           \includegraphics[clip=,width=6cm,angle=-90]{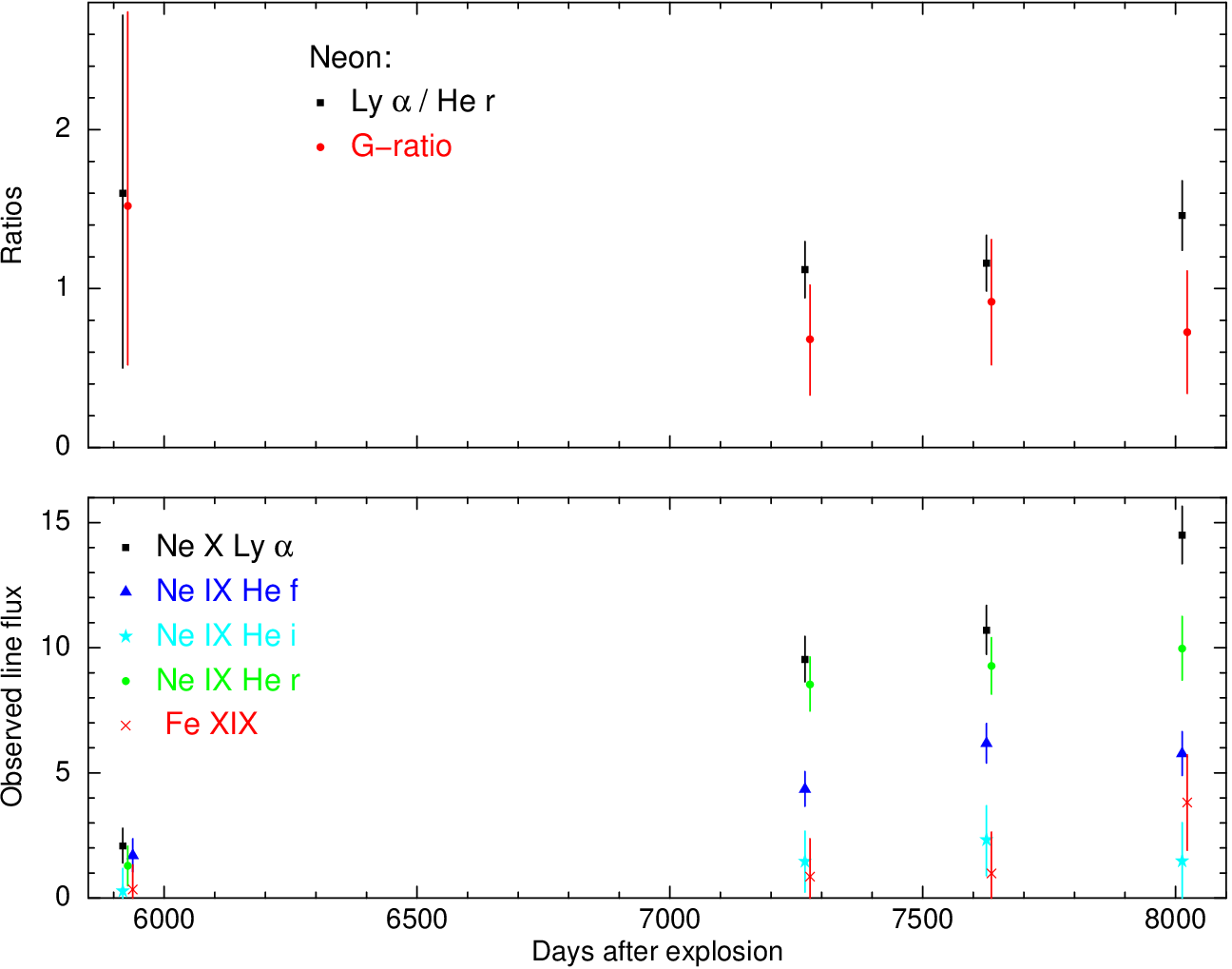}}
 \caption{The lower panels show the light curves of various emission lines with fluxes 
          in $10^{-5}$ photons cm$^{-2}$ s$^{-1}$. 
          The upper panels show the corresponding line ratios (not absorption corrected).
          The G-ratio is defined as the helium like line flux ratio of $(f+i)/r$.
          Overlapping data points are shifted by 10 days for clarity.}
\label{fig:lineratios}
\end{figure*}

The individual line fluxes are listed in Table~\ref{tab:lines}, 
light curves for prominent oxygen and neon lines are shown in Fig.~\ref{fig:lineratios}.
We note, that the temperature of the quasi-bremsstrahlung continuum shows a trend of an increase 
(254$_{-40}^{+64}$, 413$_{-21}^{+30}$, 436$_{-19}^{+26}$ and 405$_{-14}^{+17}$  for the 2003, 2007, 2008 and 2009 observation, respectively),
but stress, that this component not represents bremsstrahlung only.

%--------------------------
%--------------------------

%Fe-line in EPIC pn
We also searched for a Fe~K line complex between 6.4 and 6.7~keV in a summed EPIC-pn spectrum 
of the three observations between 2007 and 2009, where we have a better statistics in this energy band.
To characterise the feature, we fitted a bremsstrahlung continuum (kT=$(3.3\pm 0.5)$~keV, $EM=(1.3\pm0.3)\times 10^{58}$cm$^{-3}$) 
and one Gaussian in the 4.0$-$10.0~keV band. 

The best fit Gaussian centre energy is (6.57$\pm$0.08)~keV with a 
sigma-width of 62.7$_{-62.7}^{+141}$~eV and the line flux is
7.3$_{-4.2}^{+5.7} \times 10^{-7}$ photons cm$^{-2}$ s$^{-1}$.
Already, \citet{2008ApJ...676..361H} noted a possible detection in the 2007 data,
but the individual EPIC-pn and the summed EPIC-MOS spectra have not enough statistics for a detailed analysis.
With increasing ionization, the centroid energy of the Fe K line complex is shifted from the 6.4~keV fluorescent line 
over various ionization stages to the He-like emission of \ion{Fe}{xxv} at $\sim$6.7.
To account for this effect, we fitted an exposure weighted sum of the plane-parallel shock model 
derived for the high temperature component in Sect.~\ref{cap:analysis:plasma}, 
but used NEI-model version 1.1, which contains also Fe-ions less ionized than He-like.
We only allowed a constant factor to account for the normalization
and obtained a flux of the iron K lines from the shocked plasma of $4.9\times 10^{-7}$ photons cm$^{-2}$ s$^{-1}$.
As can be seen in the right panel of Fig.~\ref{fig:lines}, the resulting line shape does not 
explain the observed feature at the lower energy side.
We investigated the possibility of an additional line with zero width (also shown in Fig.~\ref{fig:lines}).
The best fit values are $4.7\pm3.3 \times 10^{-7}$ photons cm$^{-2}$ s$^{-1}$ 
at a centre energy of 6.48$_{-0.12}^{+0.17}$~keV.

\begin{table*}
\caption[]{Identified lines in the RGS spectra.}
\begin{center}
\begin{tabular}{llllrrrr}
\hline\hline\noalign{\smallskip}
\multicolumn{1}{l}{Line} &
\multicolumn{1}{l}{$E_{\rm rest}^{(a)}$~(keV)} &	
\multicolumn{1}{l}{$E_{\rm obs}^{(b)}$~(keV)} &	
\multicolumn{1}{c}{$\sigma^{(c)}$~(eV)} &	
\multicolumn{4}{c}{detected flux$^{(d)}$ ($10^{-6}$ photons cm$^{-2}$s$^{-1}$) } \\
\multicolumn{1}{l}{} &
\multicolumn{1}{l}{} &	
\multicolumn{1}{l}{} &	
\multicolumn{1}{c}{} &	
\multicolumn{1}{c}{2003} &	
\multicolumn{1}{c}{2007} &
\multicolumn{1}{c}{2008} &	
\multicolumn{1}{c}{2009} \\
\noalign{\smallskip}\hline\noalign{\smallskip}
N VI f              &	0.4198  &0.4193    & 0.01		&     $ < $33.8 &	$ < $36.7 &	   $ < $31.4 &   20.3  $\pm$8.7  \\
N VI i              &	0.4263  &0.4257    & 0.01		&     $ < $34.7 &	$ < $9.5  &	   $ < $8.7  &	  $ < $13.1      \\
N VI r              &	0.4307  &0.4301(2) & 0.01 (0$-$0.37)	&     $ < $31.4 & 24.9  $\pm$10.7 &  29.1  $\pm$9.9  &   22.8  $\pm$9.0  \\
N VI He $\beta$     &	0.4980  &0.4965(3) & 0.02 (0$-$0.74)	&7.6  $\pm$6.9  & 9.4	$\pm$4.3  &  15.4  $\pm$5.0  &   10.6  $\pm$5.2  \\
N VII Ly $\alpha$   &	0.5003  &0.4997(1) & 0.37 (0.26$-$0.46) &26.2 $\pm$7.9  & 110.9 $\pm$8.5  &  122.4 $\pm$9.0  &   134.3 $\pm$10.1 \\
O VII f	            &	0.5611  &0.5604    & 0.40		&19.9 $\pm$9.6  & 28.9  $\pm$8.6  &  37.9  $\pm$8.9  &   51.3  $\pm$9.8  \\
O VII i	            &	0.5686  &0.5680    & 0.40		&10.8 $\pm$9.0  &	$ < $15.5 &  16.1  $\pm$10.6 &   17.5  $\pm$10.1 \\
O VII r	            &	0.5739  &0.5733(1) & 0.40 (0.04$-$0.65) &13.1 $\pm$8.4  & 43.5  $\pm$8.8  &  53.1  $\pm$12.2 &   51.3  $\pm$10.0 \\
N VII Ly $\beta$    &	0.5929  &0.5926(3) & 0    (0$-$0.77)	&     $ < $13.3 & 13.4  $\pm$6.2  &  22.6  $\pm$6.4  &   23.3  $\pm$7.6  \\
N VII Ly $\gamma$   &	0.6254  &0.6247(7) & 0.63		&     $ < $9.0  &	$ < $3.7  &	   $ < $5.3  &   5.2   $\pm$4.7  \\
N VII Ly $\delta$   &	0.6404  &0.6406(8) & 0.63		&15.3 $\pm$5.2  &	$ < $1.7  &	   $ < $2.3  &   7.2   $\pm$4.6  \\
O VIII Ly $\alpha$  &	0.6537  &0.6529(1) & 0.63		&40.5 $\pm$5.8  & 130.8 $\pm$8.0  &  167.5 $\pm$8.4  &   177.8 $\pm$9.9  \\
O VII He $\beta$    &	0.6656  &0.6650(5) & 0.63 (0.46$-$0.72) &9.4  $\pm$4.9  & 14.2  $\pm$4.9  &  8.2   $\pm$4.9  &   16.3  $\pm$5.5  \\
O VII He $\gamma$   &	0.6978  &0.697(2)  & 0.11		&5.1  $\pm$4.5  &	$ < $4.1  &  6.9   $\pm$4.3  &	  $ < $2.8       \\
Fe XVIII            &	0.7035  &0.703(1)  & 0.11		&     $ < $2.2  &	$ < $8.2  &	   $ < $7.4  &   9.8   $\pm$4.4  \\
O VII He $\delta$   &	0.7127  &0.712(1)  & 0.11		&     $ < $8.8  &	$\pm$6.1  &  5.3   $\pm$4.0  &	  $ < $3.6       \\
Fe XVII             &	0.7252  &0.7250    & 0.11		&     $ < $13.7 & 54.1  $\pm$10.7 &  90.9  $\pm$12.5 &   84.9  $\pm$13.7 \\
Fe XVII             &	0.7271  &0.7269(1) & 0.11 (0$-$0.54)	&     $ < $6.0  & 29.9  $\pm$9.4  &  28.2  $\pm$10.1 &   46.1  $\pm$11.7 \\
Fe XVII             &	0.7389  &0.7380(2) & 0.25 (0$-$0.72)	&5.0  $\pm$4.4  & 34.4  $\pm$5.0  &  38.9  $\pm$5.2  &   49.2  $\pm$6.0  \\
Fe XIX              &	0.7696  &0.7658    & 0.20		&     $ < $9.1  & 5.1	$\pm$4.4  &	   $ < $9.4  &	  $ < $10.1      \\
Fe XVIII            &	0.7715  &0.7708    & 0.20		&6.7  $\pm$5.5  & 12.3  $\pm$5.2  &  24.9  $\pm$5.9  &   21.5  $\pm$6.6  \\
O VIII Ly $\beta$   &	0.7746  &0.7738(2) & 0.20 (0$-$0.86)	&     $ < $9.3  & 33.6  $\pm$5.6  &  36.2  $\pm$6.1  &   50.3  $\pm$7.1  \\
Fe XVIII            &	0.7835  &0.7817(8) & 0.25		&     $ < $9.4  & 6.7	$\pm$4.1  &  5.6   $\pm$4.3  &   7.7   $\pm$5.1  \\
Fe XVIII            &	0.7935  &0.7929(9) & 0.25		&     $ < $2.4  &	$ < $7.1  &  8.6   $\pm$4.0  &   5.0   $\pm$4.5  \\
Fe XVII             &	0.8023  &0.803(1)  & 0.25		&     $ < $3.6  & 12.6  $\pm$4.4  &  5.7   $\pm$4.7  &	  $ < $9.9       \\
Fe XVII             &	0.8124  &0.8113    & 0.25		&     $ < $5.2  & 26.6  $\pm$5.9  &  44.9  $\pm$6.4  &   45.3  $\pm$7.2  \\
O VIII Ly $\gamma$  &	0.8169  &0.8159(3) & 0.25 (0$-$0.78)	&5.1  $\pm$4.6  & 15.5  $\pm$6.1  &  19.8  $\pm$6.6  &   27.3  $\pm$7.4  \\
Fe XVII 	    &	0.8258  &0.8248(1) & 0.64 (0.23$-$0.89) &14.9 $\pm$4.5  & 82.7  $\pm$6.4  &  100.3 $\pm$6.9  &   123.5 $\pm$8.0  \\
O VIII Ly $\delta$  &	0.8365  &0.835(2)  & 1.89		&     $ < $4.9  &	$ < $4.5  &  5.2   $\pm$5.0  &   12.5  $\pm$5.6  \\
Fe XVIII	    &	0.8530  &0.8517(8) & 1.89		&     $ < $3.2  & 11.4  $\pm$4.6  &  16.5  $\pm$5.0  &   19.4  $\pm$5.7  \\
Fe XVIII	    &	0.8626  &0.862(1)  & 1.89		&     $ < $10.0 & 11.3  $\pm$5.5  &  18.3  $\pm$5.5  &   13.4  $\pm$6.0  \\
Fe XVIII	    &	0.8726  &0.8719(4) & 1.89 (1.45$-$2.35) &     $ < $9.0  & 30.7  $\pm$5.1  &  39.9  $\pm$5.5  &   54.6  $\pm$6.5  \\
Fe XVII 	    &	0.8968  &0.8957    & 0.85		&5.0  $\pm$4.7  & 13.4  $\pm$4.5  &  17.1  $\pm$4.9  &   14.5  $\pm$5.5  \\
Ne IX f 	    &	0.9050  &0.9039    & 0.85		&17.0 $\pm$6.6  & 43.5  $\pm$7.0  &  61.8  $\pm$7.9  &   57.7  $\pm$8.8  \\
Ne IX i 	    &	0.9148  &0.9136    & 0.85		&     $ < $11.8 & 14.6  $\pm$12.1 &  23.2  $\pm$13.7 &	  $ < $30.1      \\
Fe XIX  	    &	0.9171  &0.9161    & 0.85		&     $ < $13.2 &	$ < $23.7 &	   $ < $26.3 &   38.2  $\pm$19.0 \\
Ne IX r 	    &	0.9220  &0.9208(2) & 0.85 (0.25$-$1.23) &12.9 $\pm$7.8  & 85.3  $\pm$11.0 &  92.7  $\pm$11.3 &   99.7  $\pm$12.8 \\
Fe XX               &	0.9651  &0.965(1)  & 1.76		&     $ < $4.3  & 14.6  $\pm$5.8  &  20.1  $\pm$6.2  &   24.1  $\pm$6.9  \\
Fe XVII 	    &	1.0108  &1.009(1)  & 1.76		&     $ < $9.8  & 17.8  $\pm$6.7  &  28.1  $\pm$7.6  &   17.7  $\pm$8.0  \\
Ne X Ly $\alpha$    &	1.0219  &1.0205(3) & 1.76 (1.38$-$2.28) &20.8 $\pm$7.1  & 95.3  $\pm$9.2  &  107.0 $\pm$9.9  &   145.3 $\pm$11.5 \\
Ne IX He $\beta$    &	1.0740  &1.075(2)  & 3.76 (2.41$-$5.53) &9.4  $\pm$8.0  & 16.5  $\pm$7.3  &  24.9  $\pm$7.3  &   24.5  $\pm$8.0  \\
Ne IX He $\gamma$   &	1.1270  &1.124(3)  & 1.34		&     $ < $7.4  & 11.2  $\pm$5.1  &  10.5  $\pm$5.7  &   7.0   $\pm$5.9  \\
Fe XVII 	    &	1.1512  &1.15(5)   & 1.34		&     $ < $5.1  & 9.8	$\pm$3.8  &  10.4  $\pm$4.3  &   13.3  $\pm$4.7  \\
Ne X Ly $\beta$     &	1.2109  &1.210(2)  & 1.34 (0$-$3.94)	&     $ < $10.9 &	 $ < $9.4 &	   $ < $8.3  &	  $ < $8.8       \\
Mg XI f 	    &	1.3311  &1.3297    & 2.25		&9.0  $\pm$8.3  & 10.7  $\pm$5.9  &  13.9  $\pm$5.7  &   18.6  $\pm$5.9  \\
Mg XI i 	    &	1.3431  &1.3417    & 2.25		&     $ < $11.8 &	$ < $13.0 &	   $ < $8.8  &	  $ < $7.0       \\
Mg XI r 	    &	1.3522  &1.3508(9) & 2.25 (0$-$3.73)	&     $ < $15.3 & 25.9  $\pm$7.0  &  35.2  $\pm$6.7  &   44.4  $\pm$6.6  \\
Mg XII Ly $\alpha$  &	1.4726  &1.473(2)  & 4.25 (0$-$7.27)	&     $ < $15.0 & 12.7  $\pm$4.5  &  21.1  $\pm$5.4  &   26.6  $\pm$5.7  \\
Mg XI He $\beta$    &	1.5793  &1.582(6)  & 5.01 (0$-$17.1 )	&12.6 $\pm$10.0 & 5.6	$\pm$4.5  &  7.4   $\pm$5.3  &   6.2   $\pm$5.1  \\
Si XIII f	    &	1.8394  &1.840     & 5.38		&     $ < $31.0 &	$ < $20.9 &	   $ < $27.4 &	  $ < $27.9      \\
Si XIII i	    &	1.8537  &1.855     & 5.38		&     $ < $33.9 &	$ < $29.3 &	   $ < $4.96 &	  $ < $38.7      \\
Si XIII r	    &	1.8649  &1.865(4)  & 5.38 (0$-$15.8 )	&     $ < $36.3 & 34.4  $\pm$11.1 &	   $ < $4.68 &   39.8  $\pm$12.6 \\
Si XIV Ly $\alpha$  &	2.0060  &1.997(1)  & 0 (0$-$230)        &     $ < $85.6 & 22.5  $\pm$12.4 &  19.7  $\pm$19.4 &   18.1  $\pm$15.8 \\
\noalign{\smallskip}\hline\noalign{\smallskip}
\end{tabular}
\end{center}
Note: All errors are for $\Delta\chi^2=2.71$ confidence range. Parameters without errors were coupled 
to the corresponding parameter of the following line with higher energy (see text).\\
$^{(a)}$ Rest energy of identified line according to ATOMDB 1.3.1. If the line is a multiplet, we give the energy of the strongest component.\\
$^{(b)}$ Central energy of the fitted Gaussian.\\
$^{(c)}$ Best fit value of Gaussian sigma width.\\
$^{(d)}$ Detected flux for the individual observations. For fluxes consistent with 0 at $\Delta\chi^2 \le  2.71$ only the upper limit is given.\\
\label{tab:lines}
\end{table*}

\subsection{Plasma Evolution}
\label{cap:analysis:plasma}

%--------------------------------
\begin{figure*}
  \resizebox{0.98\hsize}{!}{\includegraphics[angle=-90]{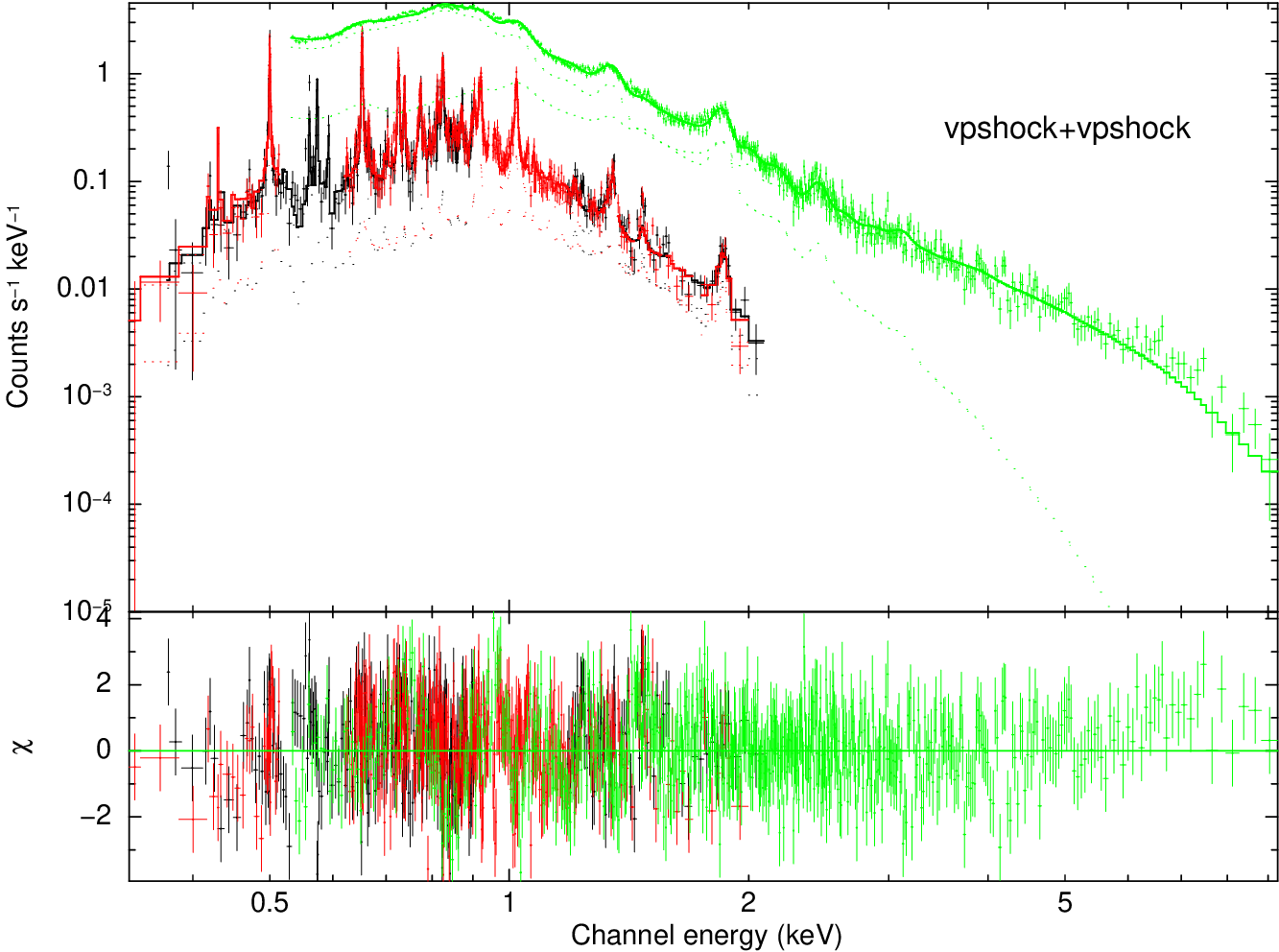}
                            \includegraphics[angle=-90]{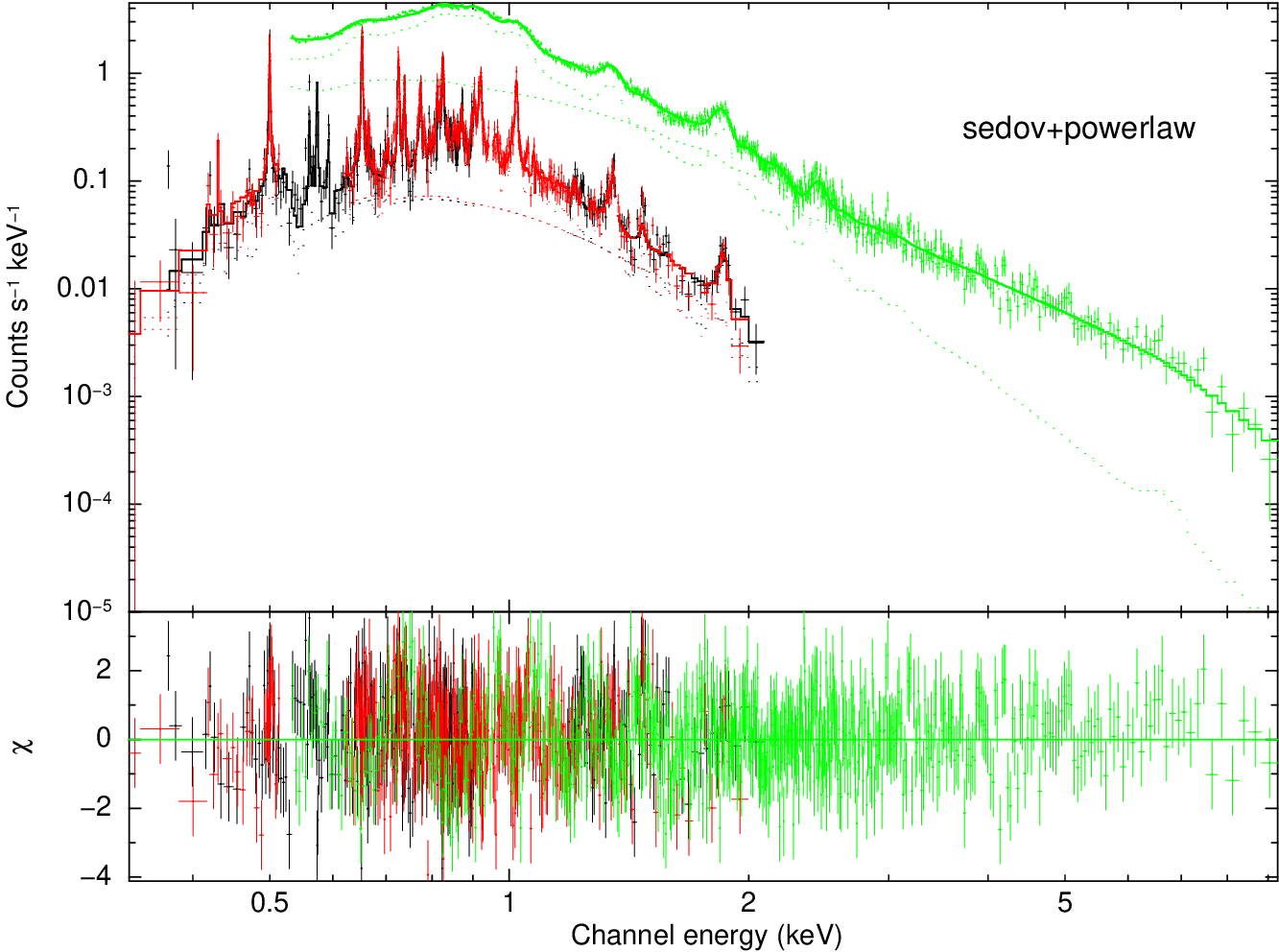}}
  \caption{The \xmm\ RGS1 (black), RGS2 (red) and EPIC-pn (green) spectra of the 2009 observation. 
           Plotted is also a model containing two plane-parallel shock components (left panel) and a sedov+powerlaw model (right panel).
          }
  \label{fig:plasmamodels}
\end{figure*}
%--------------------------------

To probe the physical parameters of the X-ray emitting plasma, we fitted plasma emission models.
Neither in \xmm\ \citep[e.g.][]{2008ApJ...676..361H}, nor in Chandra \citep[e.g.][]{2006ApJ...645..293Z} observations, 
the spectrum was described well by just one emission component. 
Zhekov et al. derived a bimodal temperature distribution in differential emission measure, peaking at $\sim$0.5~keV and $\sim$2.5~keV.
A non-thermal component might also contribute to the hard component \citep{2005ApJ...634L..73P}.
We noticed, that the emission lines in the \xmm\ spectra can also be described by the {\tt sedov}-model in XSPEC, 
which has a temperature distribution according to the Sedov self-similar solution for SNRs. 
Here a power law is necessary for the high energy tail (cf. Fig.~\ref{fig:plasmamodels} ). 

The abundances of the sedov component are significantly higher than normally found in X-ray analyses, 
thus we doubt the physical correctness of this component, 
but the model demonstrates the possible contribution of a non-thermal component 
up to $(1.44\pm0.03) \times 10^{-3}$ photons~keV$^{-1}$cm$^{-2}$s$^{-1}$ at 1~keV with power law index $\Gamma=3.00\pm0.02$ (for the 2009 observation).
Formally the {\tt sedov+powerlaw} model results in a better fit (e.g. for the 2009 observation: $\chi^{2}_{\rm red} = 1.35$ vs. $1.48$).

However, we decided to use the approximation of a plane-parallel shock structure at two temperatures as commonly found in the literature,
since the {\tt sedov} model is numerically slow and the RGS does not provide high resolved lines above 2.0~keV.

For a precise determination of the high temperature plasma component, 
we extended the energy band up to 10~keV by fitting the RGS1,2 and EPIC-pn spectra simultaneously, as it is successfully done for calibration \citep{2008SPIE.7011E..68P}.

A comparison of the empiric model fitted to the 2009 RGS data and to a dataset containing also the EPIC-pn spectrum in the 0.2$-$2.0 keV energy range shows,
that again 20 of 21 line widths were consistent. Because of the lower energy resolution of EPIC-pn the line widths are determined by the RGS spectra.
We found inconsistent fluxes of the helium-like \ion{N}{vi} and \ion{N}{vii} Ly$_\alpha$ line between the EPIC-pn and RGS spectra.
This is most likely due to calibration problems in the modeled redistribution of the EPIC-pn response at low energies. 
Thus we decided to limit the EPIC-pn band to 0.53$-$10~keV for this study.
In future studies the $0.2-0.53$~keV EPIC-pn band with advanced calibration will yield additional information.

To derive the plasma evolution we fitted a two component plane-parallel shock model \citep[{\tt vpshock},][]{2001ApJ...548..820B} 
with NEI-version 2.0 and ionization timescale range $0<\tau<\tau_{\rm u}$. 
Analogous to \citet[][and references therein]{2009ApJ...692.1190Z} the chemical composition of the plasma component is assumed to be constant in time 
and the abundances were fitted with the exception of He (set to 2.57), C (0.09), Ar (0.54), Ca (0.34) and Ni (0.62), 
which produce no significant features in our data. 
The photo-electric absorption by the Galactic interstellar medium was set to \nh\ = 6\expo{20}cm$^{-2}$, 
whereas the LMC column density with abundances set to 0.5 for metals was a free parameter.
The abundances for absorption and emission components are given according to \citet{2000ApJ...542..914W}, 
which resulted in a slightly better fit ($\chi^{2}_{\rm red} = 1.32$ vs. $1.38$) 
than the abundances of \citet{1989GeCoA..53..197A}.
The main impact is on the absorption, which influences continuum emission and line ratios.
We found no impact on the main results of our study.
A constant factor is allowed to vary for calibration differences of the individual instruments. 
We modified the shock model line widths to be described by the power law function mentioned above, thus our
custom version of the {\tt vpshock} model contains two more parameters ($\sigma_0$, $\alpha$). 

We fitted the four XMM observations (12 spectra) simultaneously, which results in a $\chi^2/{\rm dof}=5400/4084$. 
As best fit parameters we obtained for the LMC absorption 
\nh = $(2.23\pm0.02)\times 10^{21}$ cm$^{-2}$, 
for the systemic velocity 
v$_{\rm 87A}=326_{-16}^{+13}$ \kms
and for the instrument dependent constants 
c$_{\rm RGS2}=1.03\pm0.01$ and 
c$_{\rm pn}=1.04\pm0.01$ (relative to c$_{\rm RGS1}=1$). 
For the time dependent parameters see Table~\ref{tab-fit-xmm} and Fig.~\ref{fig:vpshock-evolution},
for the abundances see Table~\ref{tab:abundances}.

%-------------------------------------------------
\begin{table}
\caption[]{Results from the VPSHOCK+VPSHOCK model fits.}
\begin{center}
\begin{tabular}{lrrrr}
\hline\hline\noalign{\smallskip}
\multicolumn{1}{l}{Parameter} &
\multicolumn{1}{c}{May 2003}&
\multicolumn{1}{c}{Jan. 2007}&
\multicolumn{1}{c}{Jan. 2008}&
\multicolumn{1}{c}{Jan. 2009}\\				
\noalign{\smallskip}\hline\noalign{\smallskip}	\vspace{1mm}
$kT_{1}$(keV)		& 0.41$_{-0.01}^{+0.01}$	& 0.52$_{-0.003}^{+0.002}$	& 0.53$_{-0.001}^{+0.002}$	&0.54$_{-0.001}^{+0.002}$\\		
$kT_{2}$(keV)		& 2.89$_{-0.19}^{+0.19}$	& 2.43$_{-0.06}^{+0.06}$		& 2.50$_{-0.05}^{+0.05}$		&2.38$_{-0.05}^{+0.04}$	\\ \vspace{1mm}
$EM_{1}^{(a)}$		& 4.09$_{-0.1}^{+0.1}$	& 17.83$_{-0.12}^{+0.11}$	& 23.45$_{-0.12}^{+0.13}$	&28.37$_{-0.14}^{+0.15}$	\\ \vspace{1mm}
$EM_{2}^{(a)}$		& 1.06$_{-0.03}^{+0.03}$	& 2.95$_{-0.05}^{+0.04}$		& 3.91$_{-0.05}^{+0.06}$		&4.91$_{-0.07}^{+0.05}$	\\ \vspace{1mm}	
$\tau_{{\rm u,}1}^{(b)}$	& 3.25$_{-0.26}^{+0.29}$	& 4.82$_{-0.12}^{+0.14}$		& 6.03$_{-0.14}^{+0.19}$		&6.86$_{-0.18}^{+0.21}$	\\ \vspace{1mm}	
$\tau_{{\rm u,}2}^{(b)}$	& 1.33$_{-0.1}^{+0.12}$	& 1.93$_{-0.1}^{+0.11}$		& 2.02$_{-0.1}^{+0.09}$		&2.22$_{-0.09}^{+0.13}$	\\ \vspace{1mm}	
$\sigma_0$(eV)		& 3.51$_{-0.94}^{+1.11}$	& 1.51$_{-0.4}^{+0.25}$		& 1.58$_{-0.29}^{+0.31}$		&1.43$_{-0.29}^{+0.33}$	\\ \vspace{1mm}	
$\alpha$		& 1.10$_{-0.76}^{+0.74}$	& 2.64$_{-0.45}^{+0.69}$		& 3.15$_{-0.63}^{+0.43}$		&3.10$_{-0.52}^{+0.6}$	\\		
\noalign{\smallskip}\hline\noalign{\smallskip}
\end{tabular}
\end{center}
\label{tab-fit-xmm}
Note: Errors for $\Delta\chi^2=2.71$ confidence range.\\
$^a$ EM$ = \int n_e n_H dV$  in $10^{58}$ cm$^{-3}$, a distance of 50 kpc is assumed.\\
$^b$ $\tau_{\rm u}$: upper limit on the ionization time range in $10^{11}$ cm$^{-3}$ s.

\end{table}
%-------------------------------------------------

%-------------------------------------------------
\begin{table}
\caption[]{Abundances of SN 1987 A}
\begin{center}
\begin{tabular}{lrrrrrr}
\hline\hline\noalign{\smallskip}
\multicolumn{1}{c}{ } &
\multicolumn{1}{c}{2vpshock} &	
\multicolumn{1}{c}{sedov+po} &
\multicolumn{1}{c}{H08} &
\multicolumn{1}{c}{Z09} &	
\multicolumn{1}{c}{LF96} &
\multicolumn{1}{c}{H98} \\
\noalign{\smallskip}\hline\noalign{\smallskip}   
N&	1.385$_{-0.057}^{+0.044}$&	3.34$_{-0.19}^{+0.34}$&	0.54	&	0.83	&	2.37	&		\vspace{1mm}\\	
O&	0.128$_{-0.002}^{+0.002}$&	0.40$_{-0.01}^{+0.01}$&	0.05	&	0.14	&	0.33	&	0.33	\vspace{1mm}\\	
Ne&	0.338$_{-0.006}^{+0.004}$&	1.00$_{-0.03}^{+0.03}$&	0.20	&	0.41	&	0.63	&	0.41	\vspace{1mm}\\	
Mg&	0.291$_{-0.007}^{+0.008}$&	0.71$_{-0.03}^{+0.04}$&	0.17	&	0.42	&		&	0.48	\vspace{1mm}\\	
Si&	0.516$_{-0.016}^{+0.018}$&	1.01$_{-0.04}^{+0.08}$&	0.48	&	0.63	&	0.91	&	0.59	\vspace{1mm}\\	
S &	0.451$_{-0.041}^{+0.039}$&	1.56$_{-0.19}^{+0.29}$&	0.59	&	0.40	&	0.46	&	0.48	\vspace{1mm}\\	
Fe&	0.224$_{-0.002}^{+0.003}$&	0.46$_{-0.01}^{+0.10}$&	0.09	&	0.33	&	1.12	&	0.38	\\

\noalign{\smallskip}\hline\noalign{\smallskip}
\end{tabular}
\end{center}
Note: Abundances relative to \citet{2000ApJ...542..914W}.
2vpshock: to component plane parallel shock model fitted simultaneously to all 12 spectra (this study),
sedov+po: sedov model with non thermal component fitted to the 2009 data (this study),
H08:  EPIC-pn 2007 \citep{2008ApJ...676..361H} VPSHOCK+VPSHOCK W00 model,
H09:  Chandra results \citep{2009ApJ...692.1190Z},
LF96: Inner ring \citep{1996ApJ...464..924L},
H98:  LMC-average \citep{1998ApJ...505..732H}.
\label{tab:abundances}
\end{table}
%-------------------------------------------------

\subsection{Integrated Flux}

To derive the detected flux analogous to \citet{2006A&A...460..811H} and \citet{2008ApJ...676..361H}, 
the vpshock+vpshock model was fitted separately to the most recent two EPIC-pn spectra. 
The fluxes do not depend strongly on individual model parameters. 
The fluxes for various sub-bands are given in Table~\ref{tab:flux}.

%-------------------------------------------------
\begin{table}
\caption[]{EPIC-pn fluxes}
\begin{center}
\begin{tabular}{crr}
\hline\hline\noalign{\smallskip}
\multicolumn{1}{c}{sub-band} &	
\multicolumn{1}{c}{Rev. 1482} &
\multicolumn{1}{c}{Rev. 1675} \\
\multicolumn{1}{c}{(keV)} &	
\multicolumn{1}{c}{Jan 2008} &
\multicolumn{1}{c}{Jan 2009} \\
\noalign{\smallskip}\hline\noalign{\smallskip}   
0.2$-$0.8   &	1.38$_{-0.01}^{+0.01}$&	1.59$_{-0.01}^{+0.01}$\vspace{1mm}\\	
0.8$-$1.2   &	2.04$_{-0.02}^{+0.02}$&	2.48$_{-0.02}^{+0.01}$\vspace{1mm}\\	
1.2$-$8.0   &	2.06$_{-0.05}^{+0.05}$&	2.53$_{-0.04}^{+0.05}$\vspace{1mm}\\	
0.5$-$2.0   &	4.31$_{-0.03}^{+0.03}$&	5.24$_{-0.04}^{+0.04}$\vspace{1mm}\\	
3.0$-$10.0  &	0.51$_{-0.02}^{+0.01}$&	0.58$_{-0.02}^{+0.02}$\vspace{1mm}\\	
0.5$-$10.0  &	5.28$_{-0.08}^{+0.05}$&	6.40$_{-0.08}^{+0.05}$\vspace{1mm}\\	
0.2$-$10.0  &	5.51$_{-0.07}^{+0.06}$&	6.64$_{-0.07}^{+0.07}$\\	
\noalign{\smallskip}\hline\noalign{\smallskip}
\end{tabular}
\end{center}
Note: Fluxes in \oergcm{-12}. Errors are for 68\% confidence.
For the fluxes of the previous observations see \citet{2006A&A...460..811H} and \citet{2008ApJ...676..361H}.
\label{tab:flux}
\end{table}
%-------------------------------------------------

\section{Discussion}

%--------------------------
\begin{figure*}
  \resizebox{.95\hsize}{!}{\includegraphics[clip=,width=2.4cm,angle=-90]{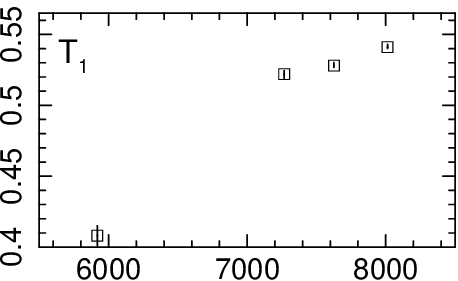}
                           \includegraphics[clip=,width=2.4cm,angle=-90]{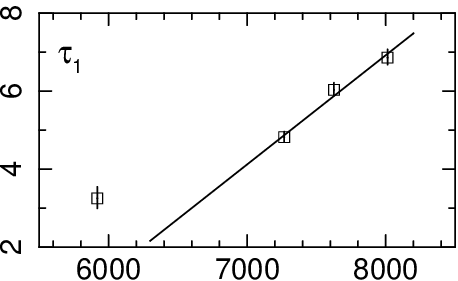}
                           \includegraphics[clip=,width=2.4cm,angle=-90]{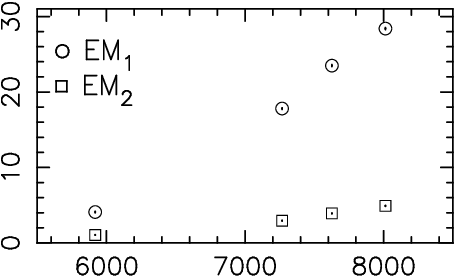}
                           \includegraphics[clip=,width=2.4cm,angle=-90]{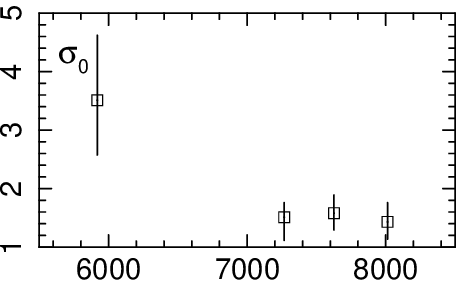}}
\newline
  \resizebox{.95\hsize}{!}{\includegraphics[clip=,width=2.4cm,angle=-90]{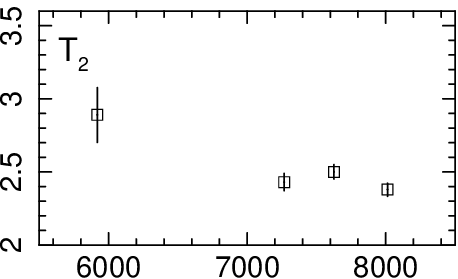}
                           \includegraphics[clip=,width=2.4cm,angle=-90]{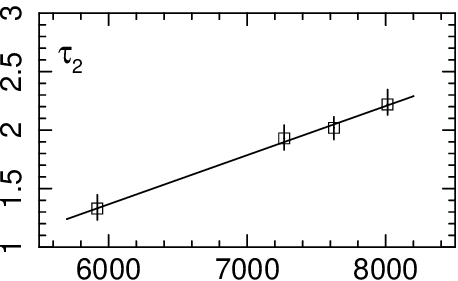}
                           \includegraphics[clip=,width=2.4cm,angle=-90]{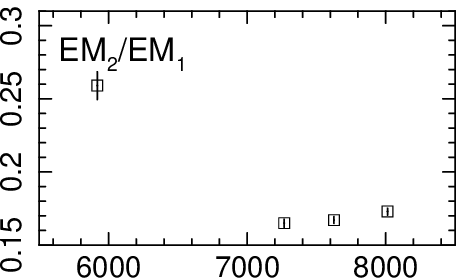}
                           \includegraphics[clip=,width=2.4cm,angle=-90]{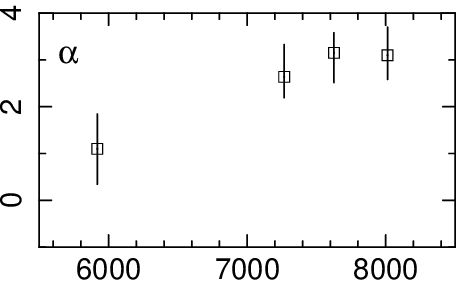}}
 \caption{Time evolution of the plasma variables: Horizontal axes are days after explosion. 
          $T_{1,2}$ is given in keV, EM$_{1,2}$ in 10$^{58}$cm$^{-3}$, $\tau_{1,2}$ in $10^{11}$cm$^{-3}$s and $\sigma_0$ in eV. }
\label{fig:vpshock-evolution}
\end{figure*}
%--------------------------

%--------------------------
\begin{figure}
  \resizebox{\hsize}{!}{\includegraphics[angle=-90]{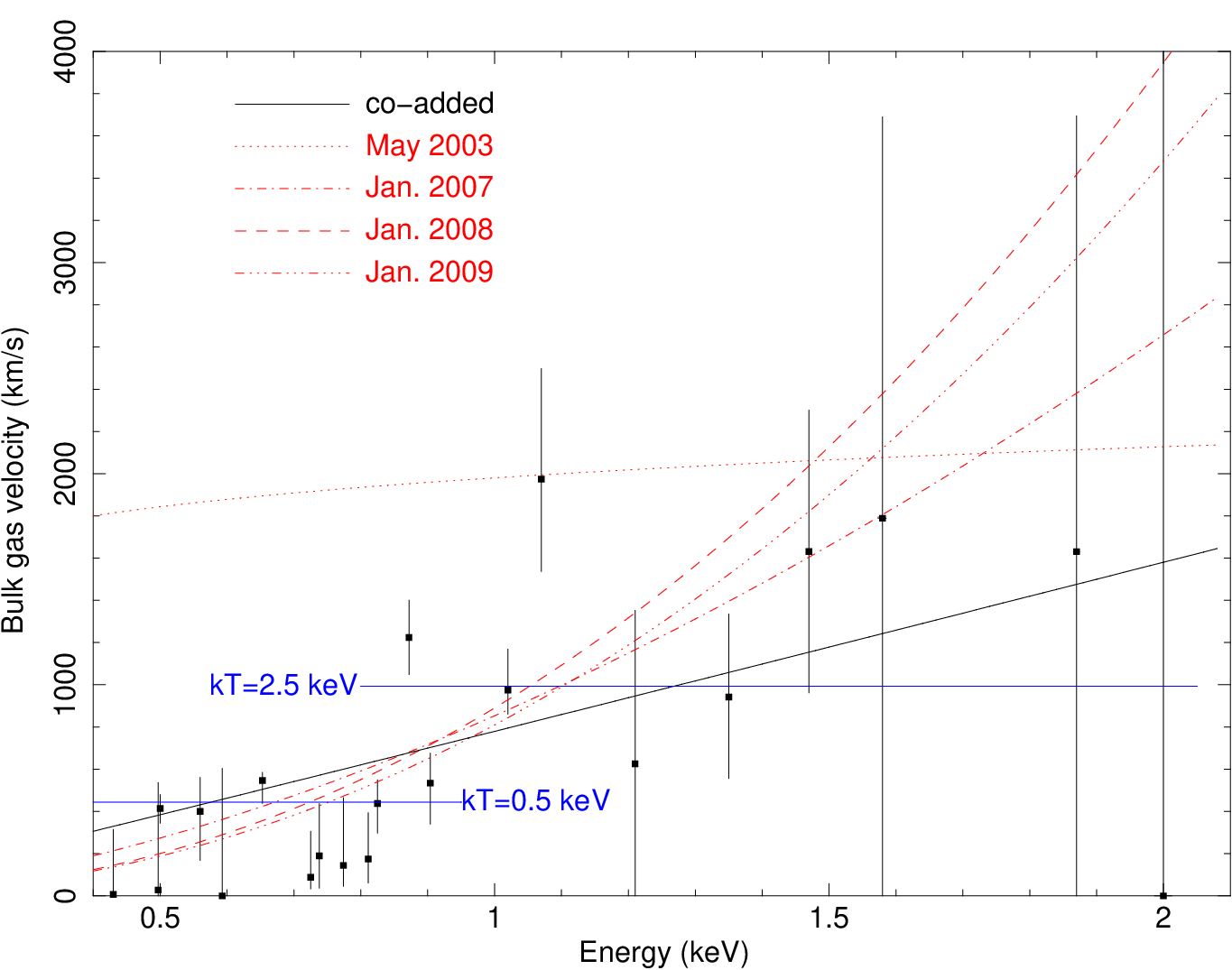}}
  \caption{Squares denote the bulk gas velocity inferred from the emission line broadening 
           derived from the Gaussian model fitted to the co-added RGS spectra. 
           Errors are for $\Delta\chi^2=1$. 
           The solid black line shows the best fit power law function to these data points. 
           The other (red) lines show the best fit power law functions 
           as derived with the modified vpshock model from the individual spectra.
           For clarity, the errors are not plotted. Note, that especially for the 2003 data the errors are large (cf. Fig.~\ref{fig:vpshock-evolution} right panels).
           The horizontal blue lines indicate the post shock gas velocities resultant from strong adiabatic shocks 
           that heat the plasma to a temperatures of $kT= 0.5$ and $2.5$~keV after thermal equilibration.}
  \label{fig:linewidth}
\end{figure}
%--------------------------

We monitored \SN\ with \xmm\ and obtained RGS spectra with unprecedented statistical quality. 
We analyzed individual line shapes and fluxes with an empirical model and applied 
physical models to a combined set of RGS and EPIC-pn spectra.

The line shifts derived from the empirical model ($349\pm24$ \kms) and the plasma codes ($326\pm16$ \kms) are consistent.
Considering the RGS wavelength scale calibration the line shifts are consistent with 
the systemic velocity derived from optical \sn\ observations \citep[286.74$\pm$0.05~\kms, e.g.][]{2008A&A...492..481G}.

%Line broadening
Line broadening is significantly detected in the RGS spectra. 
To obtain a conversion factor for the measured line widths and the bulk gas velocity, 
we convolved the emission spectrum of a cylindrically expanding ring 
at an inclination of 45\degr and a temperature of $ kT = 0.5$~keV 
in a bulk gas velocity range of 0 $<v_{\rm bulk} < 1000$~\kms\ with the RGS response 
and fitted a Gaussian to the resultant line profile. 
In this simulation we get v$_{\rm bulk}=0.80 \times \Delta$v (FWHM). 
For a similar analysis see \citet[][]{2002ApJ...574..166M}.

As seen in Fig.~\ref{fig:linewidth}, v$_{\rm bulk}$ is determined best by the profiles of the two strongest lines 
\ion{N}{vii} Ly$_{\alpha}$ and \ion{O}{viii} Ly$_{\alpha}$.
The broadening of the former line including a statistical error of $\Delta\chi^2=2.71$ is $518_{-150}^{+130}$ \kms\ FWHM.
Since the instrumental line broadening is somewhat higher, 
the uncertainty of the RGS line spread function (lsf) causes a similar error. 
A 10\% uncertainty of the lsf ($\Delta \lambda_{\rm lsf} \sim 0.07 \AA\ $ FWHM) 
gives an uncertainty range of 400$-$600 \kms\ FWHM. 
But recent studies indicate that the current lsf-model rather 
over-estimates the instrumental line width (Jelle Kaastra, private communication). 
For our modeling of the RGS lines the extent of the source ($\sim$$1\farcs5$) is negligible. 
For a thermally equilibrated plasma, as it is expected for the shock velocity range of the lower temperature component \citep{2005AdSpR..35.1017R},
the thermal line broadening also does not contribute significantly (e.g. 133 \kms\ FWHM for nitrogen at $kT=0.5$~keV).
We also found no influence due to the two unresolved Lyman transitions by modeling the line with two Gaussians, 
and other unresolved lines should not contribute significantly either. 

For the \ion{N}{vii} Ly$_{\alpha}$ line this indicates a bulk velocity of $(414\pm120)$~\kms. 
In the case of gas shocked by a strong adiabatic shock wave and thermal equilibration, 
this yields v$_{\rm sh}= 4/3 {\rm v}_{\rm bulk}  = 552$ \kms and a post shock temperature of $ kT= 3/16 \mu {\rm v}_{\rm sh}^2 =0.435_{-0.22}^{+0.25}$ keV. 
Here an adiabatic index of $\gamma = 5/3$ and an average molecular weight of $\mu = 0.73m_{\rm p}$ 
according to the abundances of the inner ring is assumed.
Similarly for the \ion{O}{viii} Ly$_{\alpha}$ line we obtain $ kT=0.75_{-0.36}^{+0.22}$ keV.
Thus these line widths are consistent with the temperatures derived from the plasma models.

The bulk gas velocities derived from these two lines are also consistent with the result of the Chandra LETG observation in 2007 
\citep[$\sim$360~\kms, ][]{2009ApJ...692.1190Z}.
But in general the bulk gas velocities as function of line energy which can be described globaly by power laws, 
are higher for the RGS derived values than the values deduced from the Chandra spectra (e.g. by a factor of $\sim$2 at 1~keV).
Some of this difference is caused by thermal broadening, which contribute to the \xmm\ derived line widths, 
but not to the Chandra values, which are based on spatial spectral deformation 
(i.e. the bulk gas velocity is deduced from the difference of the line broadening in the two grating arms).
But this effect would only become significant for plasma out of thermal equilibrium.
Also the radial velocity distribution of the shocked plasma might contribute differently to the line broadening 
seen in the RGS and dispersed Chandra spectra, but it can not explain such a large difference.
We believe that most of the difference is caused by the different modeling methods and by calibration differences.

%Line broadening - evolution
With our modified plasma emission code we can also investigate the evolution of the line broadening 
and we can narrow the artificial broadening due to blended lines. 
The increase in $\alpha$ indicates that the velocity of regions with lower temperature 
decreases relative to the regions with higher temperature. 
The RGS derived line widths are determined most accurately in the 0.5-1.0~keV  band, 
where we have the best statistics and strong emission lines.
Conspicuous is the sudden decrease of the line broadening in this energy band between 2003 and 2007,
indicating a deceleration of the bulk velocity.
Although the statistics of this observation is low, 
a comparison with a line broadening fixed to the function derived from the empirical model shows 
that the individually derived line description is significant (f-test probability $1.4\times10^{-4}$).
From 2007 to 2009 the \xmm\ derived line widths show a rather constant line width, 
similar to those of the Chandra grating observations in 2004 and 2007. 
But from the HETG observation in 1999 \citet[][]{2002ApJ...574..166M} report line widths corresponding to a post shock plasma velocity of $\sim$2500 \kms.
Thus the line broadening had to decrease in between the \xmm\ observation (5918 SN days) and 
the Chandra LETG sequence ($\sim$6400 SN days) by $\sim$50\% at 1~keV.
We note, that the sudden deceleration of the expansion seen in the X-ray images around day 6100 
\citep{2009ApJ...703.1752R} also falls into this time interval.
But we emphasize that the spectra in 2003 might be more complex than assumed by two shock components 
and could contain a more complex mixture of radial velocity and temperature components.

%--------------PLASMA PARAMETERS
Also the derived plasma parameters of the plasma emission model show a clear evolution between 2003 and 2007. 
%EM
The emission measure ratio $EM_2/EM_1$ shows a decrease.
During this time also the soft flux was rising drastically. 
In the later observations the $EM$ ratio is rather constant with a tendency of an increase.

%ionization times:
We also observe the ongoing ionization of the plasma.
This causes e.g. the increase of the [\ion{Ne}{x} Ly$_{\alpha}$]/[\ion{Ne}{ix} He r] 
flux ratio by $\sim$26\% during the last two observations.
The upper limit of the ionization timescale range $\tau_{\rm u}$ is steadily increasing for both plasma components.
$\tau_{\rm u,1}$ indicates some upturn after the first observation, which could be caused by an increasing density.
A fit of a linear function to the last three values (see Fig.~\ref{fig:vpshock-evolution}) yields an ''average'' density of $n_1= (3.24\pm0.64)$\expo{3} cm$^{-3}$.
For the high temperature component no evolution in density is observed and a similar fit yields $n_2= (4.8\pm1.4)$\expo{2} cm$^{-3}$.
The ratio of the emission measure of the low and high temperature components is roughly constant with a volume ratio of $\sim$7.5.

The densities derived from the unshocked gas in the optical are $6\times10^{3}$ cm$^{-3}<n<3\times10^{4}$ cm$^{-3}$ for the ring 
and $\sim$$10^{2}$ cm$^{-3}$ for the extended nebula \citep{1996ApJ...464..924L}.
A strong adiabatic shock wave increases these densities by a factor of four.  
Our values would suggest less denser regions in the ring and the surrounding nebula as origin of the X-rays.
But the {\tt vpshock} model assumes a ionization time distribution linear in emission measure, 
which is clearly not true for \SN. A complex density distribution and effects of radial expansion affect the derived densities.
However, these values are not consistent with those from the modeling of the light curve \citep[][]{2006A&A...460..811H,2007AIPC..937...33A},
and should be regarded with some care.

%TEMP
The temperature of the soft plasma component shows a strong increase between the first two observations, 
and a slight increase in the later ones. 
Also from Chandra observations a rising temperature of this component is derived \citep{2006ApJ...646.1001P}.
Between 2004 and 2007 \citet{2009ApJ...692.1190Z} report a nearly constant temperature.
With RGS we can see an ongoing small increase in the temperature of the soft component. 
However, a decrease caused by the deceleration of the shock wave and adiabatic expansion of the shocked plasma 
is expected which is seen in the high temperature component. 
An increase is possible, if regions with slightly higher temperature contribute more to the emission measure with time, 
e.g. due to more rapidly rising volume of these regions.
But also the effects of incomplete thermal equilibration might contribute to a rising electron temperature, 
especially initially after the shock wave has reached denser protrusions of the ring.

%abundances
The abundances derived from the two component plan-parallel shock model (cf. Table~\ref{tab:abundances}) are higher (on average by a factor of 1.8), 
than derived by \citet[][VPSHOCK+VPSHOCK W00 model]{2008ApJ...676..361H}. Heng et al. used a similar model, 
but fitted only the EPIC-pn spectum of the 2007 observation and derived 
different plasma parameters (e.g. temperatures) which influences the derived abundances.
Compared with the Chandra grating results of \citet{2009ApJ...692.1190Z}, our abundances are consistent within 30\%, 
except nitrogen, where our value is higher (by a factor of 1.7) but rather matches the \citet{1996ApJ...464..924L} value,
and Mg and Fe, which are $\sim$30\% lower.
We see, that the individual modelling causes systematic differences, much larger than the statistical errors.
We also note, that the abundances derived from the plane-parallel shock model are lower than the abundances of the inner ring 
as derived from \citet{1996ApJ...464..924L}, whereas the abundances of the sedov+powerlaw model are higher. 
Thus the lower abundances derived so far in X-ray analysis may be caused by the assumption of a two component plan-parallel shock structure.

%Fe K
As shown in the right panel of Fig.~\ref{fig:lines}, 
we see a clear indication for an Fe~K feature.
The black curve in Fig.~\ref{fig:lines} shows the emission from the shocked plasma of the inner ring,
as expected from our two component shock model, which describes the Fe~L lines well.
Surprisingly we find, that the feature suggests a contribution of emission at lower energy 
of less ionized, or even neutral, iron as it is present in the unshocked part of the ring and in the supernova debris.
In the case of a very young ($\tau_{\rm u}\le3\times 10^{10}$) reverse shock in a iron rich region, 
we would also expect emission from other elements in the RGS spectra.
If the additional emission is caused by fluorescence (6.4~keV), this would need neutral iron.
In this case, the iron could be excited by X-rays from the shocked ring,
but this would need a much higher column density of the unshocked region.
Thus the emission might originate from the supernova debris, 
where a higher column density is possible and
reprocessed radiation from nuclear decays might contribute.

\section{Conclusions}

\begin{enumerate}
\item With our monitoring we can follow the detailed evolution of the supernova remnant of \SN. 
      E.g. the upturn in ionization age and emission measure ratio between 2003 and 2007 shows that the blast wave was propagating into the inner ring.
 
\item The decreasing line widths at lower energies in between the first two observations indicate a deceleration of the lower temperature plasma, 
      that correlates well with the decelerating ring expansion as observed in the Chandra images.

\item The electron temperature derived for the soft temperature component with plasma models is consistent 
      with the line widths of the emission lines in the corresponding energy range. 
      This is expected, if the emission is primary caused by shocks transmitted into denser regions.

\item The lower statistics at higher energies do not justify such conclusions for the high temperature component, 
      where the bulk velocity can be reduced by the contribution of reflected shocks which would also heat the plasma further.
      But the decreasing temperature with time and the higher line widths seen by \xmm\ and Chandra rather suggest forward shocks 
      in less denser regions as the dominating process.

\item The iron K feature argues rather for a thermal high energy component and little if any non-thermal contribution. 
      The line shape further suggests the contribution ($\sim$50\%) of a cold iron line (2.3$\sigma$ level), 
      possibly emitted from the supernova debris.
\end{enumerate}

Further monitoring will yield information on the arising SNR
of \SN\ and the increasing statistics of the brightening source will allow 
even more detailed analyses also beyond the assumption of plane-parallel geometry.
This might put further constraints on the origin of the X-ray emitting regions and their dynamics. 
Similarly the iron K emission can yield information on the evolution of the debris.

\bibliographystyle{aa}
\bibliography{aa_xmm_sn1987a}

\end{document}